\documentclass[twoside,twocolumn,9pt]{article}
\usepackage{extsizes}
\usepackage[super,sort&compress,comma]{natbib} 
\usepackage[version=3]{mhchem}
\usepackage[left=1.5cm, right=1.5cm, top=1.785cm, bottom=2.0cm]{geometry}
\usepackage{balance}
\usepackage{mathptmx}
\usepackage{sectsty}
\usepackage{graphicx} 
\usepackage{lastpage}
\usepackage[format=plain,justification=justified,singlelinecheck=false,font={stretch=1.125,small,sf},labelfont=bf,labelsep=space]{caption}
\usepackage{float}
\usepackage{fancyhdr}
\usepackage{fnpos}
\usepackage[english]{babel}
\addto{\captionsenglish}{%
  
}
\usepackage{array}
\usepackage{droidsans}
\usepackage{charter}
\usepackage[T1]{fontenc}
\usepackage[usenames,dvipsnames]{xcolor}
\usepackage{setspace}
\usepackage[compact]{titlesec}
\usepackage{hyperref}

\usepackage{epstopdf}

\definecolor{cream}{RGB}{222,217,201}

\begin{document}

\pagestyle{fancy}
\thispagestyle{plain}
\fancypagestyle{plain}{
\renewcommand{\headrulewidth}{0pt}
}

\makeFNbottom
\makeatletter
\renewcommand\LARGE{\@setfontsize\LARGE{15pt}{17}}
\renewcommand\Large{\@setfontsize\Large{12pt}{14}}
\renewcommand\large{\@setfontsize\large{10pt}{12}}
\renewcommand\footnotesize{\@setfontsize\footnotesize{7pt}{10}}
\makeatother

\renewcommand{\thefootnote}{\fnsymbol{footnote}}
\renewcommand\footnoterule{\vspace*{1pt}%
\color{cream}\hrule width 3.5in height 0.4pt \color{black}\vspace*{5pt}} 
\setcounter{secnumdepth}{5}

\makeatletter 
\renewcommand\@biblabel[1]{#1}            
\renewcommand\@makefntext[1]%
{\noindent\makebox[0pt][r]{\@thefnmark\,}#1}
\makeatother 
\renewcommand{\figurename}{\small{Fig.}~}
\sectionfont{\sffamily\Large}
\subsectionfont{\normalsize}
\subsubsectionfont{\bf}
\setstretch{1.125} 
\setlength{\skip\footins}{0.8cm}
\setlength{\footnotesep}{0.25cm}
\setlength{\jot}{10pt}
\titlespacing*{\section}{0pt}{4pt}{4pt}
\titlespacing*{\subsection}{0pt}{15pt}{1pt}

\fancyfoot{}
\fancyfoot[LO,RE]{\vspace{-7.1pt}\includegraphics[height=9pt]{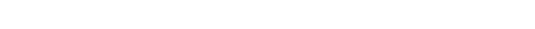}}
\fancyfoot[CO]{\vspace{-7.1pt}\hspace{13.2cm}\includegraphics{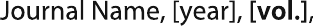}}
\fancyfoot[CE]{\vspace{-7.2pt}\hspace{-14.2cm}\includegraphics{head_foot/RF}}
\fancyfoot[RO]{\footnotesize{\sffamily{1--\pageref{LastPage} ~\textbar  \hspace{2pt}\thepage}}}
\fancyfoot[LE]{\footnotesize{\sffamily{\thepage~\textbar\hspace{3.45cm} 1--\pageref{LastPage}}}}
\fancyhead{}
\renewcommand{\headrulewidth}{0pt} 
\renewcommand{\footrulewidth}{0pt}
\setlength{\arrayrulewidth}{1pt}
\setlength{\columnsep}{6.5mm}
\setlength\bibsep{1pt}

\makeatletter 
\newlength{\figrulesep} 
\setlength{\figrulesep}{0.5\textfloatsep} 

\newcommand{\topfigrule}{\vspace*{-1pt}%
\noindent{\color{cream}\rule[-\figrulesep]{\columnwidth}{1.5pt}} }

\newcommand{\botfigrule}{\vspace*{-2pt}%
\noindent{\color{cream}\rule[\figrulesep]{\columnwidth}{1.5pt}} }

\newcommand{\dblfigrule}{\vspace*{-1pt}%
\noindent{\color{cream}\rule[-\figrulesep]{\textwidth}{1.5pt}} }

\makeatother

\twocolumn[
  \begin{@twocolumnfalse}
{\includegraphics[height=30pt]{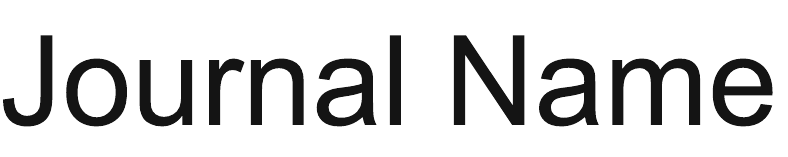}\hfill\raisebox{0pt}[0pt][0pt]{\includegraphics[height=55pt]{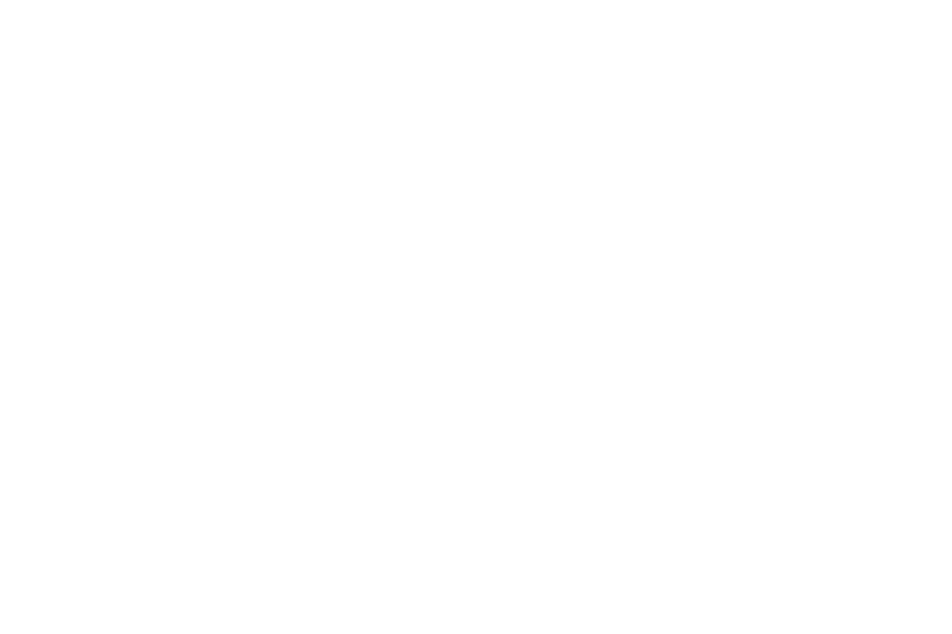}}\\[1ex]
\includegraphics[width=18.5cm]{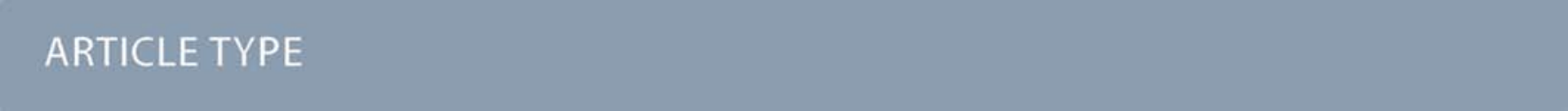}}\par
\vspace{1em}
\sffamily
\begin{tabular}{m{4.5cm} p{13.5cm} }

\includegraphics{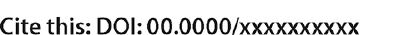} & \noindent\LARGE{\textbf{Regulating Aggregation of Colloidal Particles in an Electro-Osmotic Micropump$^\dag$}} \\
\vspace{0.3cm} & \vspace{0.3cm} \\

 & \noindent\large{Zhu Zhang,$^{\ast}$\textit{$^{a}$} Joost de Graaf,\textit{$^{b}$} and Sanli Faez\textit{$^{a}$}} \\

\includegraphics{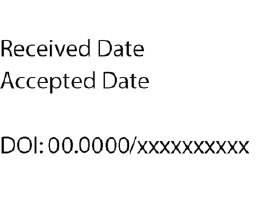} & \noindent\normalsize{Unrestricted particle transport through microfluidic channels is of paramount importance to a wide range of applications, including lab-on-a-chip devices. 
In this article, we study using video microscopy the electro-osmotic aggregation of colloidal particles at the opening of a micrometer-sized silica channel in presence of a salt gradient. 
Particle aggregation eventually leads to clogging of the channel, which may be undone by a time-adjusted reversal of the applied electric potential.
We numerically model our system via the Stokes-Poisson-Nernst-Planck equations in a geometry that approximates the real sample. 
This allows us to identify the transport processes induced by the electric field and salt gradient and to provide evidence that a balance thereof leads to aggregation. 
We further demonstrate experimentally that a net flow of colloids through the channel may be achieved by applying a square-waveform electric potential with an appropriately tuned duty cycle. 
Our results serve to guide the design of microfluidic and nanofluidic pumps that allow for controlled particle transport and provide new insights for anti-fouling in ultra-filtration.} 

\\

\end{tabular}

 \end{@twocolumnfalse} \vspace{0.6cm}
]

\renewcommand*\rmdefault{bch}\normalfont\upshape
\rmfamily
\section*{Introduction}


\footnotetext{\textit{$^{a}$~Nanophotonics, Debye Institute for Nanomaterials Science, Utrecht University, Princetonplein 1, 3584 CC Utrecht, The Netherlands, E-mail:z.zhang@uu.nl}}
\footnotetext{\textit{$^{b}$~Institute for Theoretical Physics, Center for Extreme Matter and Emergent Phenomena, Utrecht University, Princetonplein 5, 3584 CC Utrecht, The Netherlands.}}

\footnotetext{\dag~Electronic Supplementary Information (ESI) available, which consists of supporting measurement plots and three movies that show the dynamics in our system. See DOI: 00.0000/00000000.
}




Micropores and microporous structures are omnipresent building blocks of applied fluid mechanics and systems for industrial fluid processing~\cite{CHOWDHURY2019265}. 
Introducing colloidal suspensions into such structures introduces an extra level of complexity, but also is of high technological relevance in filtration, separation, or assisted concentration of colloids~\cite{driscollLeveragingCollectiveEffects2019}. 
Examples of this can be found in fields ranging from oil recovery~\cite{akbariGrowthAttachmentFacilitatedEntry2016} to food processing~\cite{nathRevisitingRecentApplications2018}. 
The interaction between channel walls and the (often charged) colloidal particles (CPs) initiates a range of multi-scale processes~\cite{sparreboomPrinciplesApplicationsNanofluidic2009}, complicating the rational design of fluid processing and lab-on-chip devices. 

Electro-osmotic pumping~(EOPing) is a common technique for circulating fluids in microfluidic channels~\cite{wang_electroosmotic_2009, peng_electroosmotic_2016, han_optoelectrofluidic_2016, LI20191}. It is the method of choice~\cite{ramosAcElectrokineticsReview1998, sparreboomPrinciplesApplicationsNanofluidic2009} for nanofluidic channels with a cross section smaller than $1~\mu\mathrm{m}^2$. 
The reason for this is that in EOPs the average flow velocity is independent of the channel diameter~\cite{hunterFoundationsColloidScience2001, Tallarek_2000}, unlike pressure-driven flow, where the average flow velocity decreases with the square of the channel diameter. 
EOP-based nanoporous membranes have been applied in various methods of analytical and physical chemistry, such as: liquid-chromatography separation~\cite{Xiong_yuan_Anvanced_2017}, femto-liter pipettes~\cite{Apeng_Femto_liter_2017}, micro-energy systems~\cite{Kilsung_micropower_2012}, and micro-nanofluidic microchips~\cite{yangLowvoltageEfficientElectroosmotic2019}. 
A common challenge facing the field is that in many situations involving EOPs, there is a tendency for the microfluidic device to become clogged with aggregates of CPs. 
This limits the effectiveness these setups and their durability, and presents a major obstacle to their technological application.

Here, we study the transport of CPs through a straight silica microchannel by EOP, and its potential clogging, using video microscopy and numerical finite-element-based modeling. 
We imposed a time-dependent electric field and a salt gradient, and investigated the combined effect thereof on the behavior of the CPs. 
Experimentally, we observed CP aggregation at one end of the channel when the imposed electric field is in the direction opposite to the salt gradient, leading to eventual clogging. 
We have investigated the underlying physical processes to this phenomenon numerically by solving for the fluid flow, electric field, and salt concentration profiles in a representative geometry, as specified by the Stokes-Poisson-Nernst-Planck (SPNP) equations. 
Our model shows that aggregation results from a balance between fluid advection and particle electro-phoresis, both of which are modulated strongly by the presence of a salt gradient. 
This balance leads to fluid vortices forming at the channel opening, which are representative of the flow patterns observed in the experiment and which initiate the accumulation of CPs.

Experimentally, we further investigated the conditions for preventing the irreversible clogging of the channel. 
The aggregated CPs could be rapidly cleared from the opening by inverting the applied potential for a short period of time; as much as an order of magnitude shorter than the period associated with clogging. 
This observation enabled us to formulate a procedure to avoid clogging through the application of an asymmetric square waveform potential with appropriately tuned duty cycle and subsequently demonstrate its efficacy. 
We showed this time-dependent potential can indeed result in a net CP flow through the channel. 
This insight, combined with our understanding of the underlying interplay between electric-field- and fluid-driven transport processes will prove valuable for the future design of non-clogging EOPs for use in,~\textit{e.g.}, lab-on-a-chip devices, microfluidic nanoparticle manipulation, and industrial processes such as ultrafiltration.

\section*{Electrically Driven Colloid Transport}

In this section, we place our work into context by briefly reviewing recent work on the electrically driven transport of CPs through micro- and nanofluidic channels. The two dominant effects are: (i) fluid flow through the channel, be that pressure driven or osmotic, and (ii) phoretic transport of CPs. A typical feature of transport is the dominance of one of these over the other, while a balance --- opposing with comparable magnitude --- of these two can lead to CP-aggregation and eventual clogging.

A common way of inducing flow through a charged nano- or micropore containing an electrolyte is to impose an electric field (E-field) over the pore~\cite{LI20191}. 
The E-field acts on the charge excess in the double layer screening the charge on the surface of the pore. 
As a result, an electro-osmotic flow of the fluid is induced in the channel,~\textit{i.e.}, an EOP is formed. 
Such a pump is made more effective by the dielectric contrast between the fluid and the (insulating) material of the pore, which causes the E-field to decay predominantly through the pore~\cite{yangElectroosmoticFlowMicrochannels2001}. 
That is, the density of field lines that go through the material of the pore is negligible compared to that going through the pore opening. 
This in turn leads to a strong electro-osmotic pumping effect.

A change in dielectric properties and the associated gradients in the E-field can also be exploited to trap/separate particles.
For example, microfabricated post, ridges, and hurdles have been used in this manner~\cite{Cummings_Diele_2003, Lapizco-Encinas_dielec_2004, Barrett_Dielec_2005, Baylon-Cardiel_Diele_2010}. 
Zhu~\textit{et al.} similarly observed focusing and separation processes of microparticles and cells in spiral and serpentine microchannels~\cite{zhu_dc_2009,zhu_continuous_2010, zhu_curvature-induced_2011, zhu_continuous-flow_2011}. 
Dielectric differences can further lead to induced-charge electro-osmosis (ICEO) --- a non-linear electro-osmotic slip velocity that occurs when an applied E-field acts on the ionic charge it induces around a polarizable (potentially uncharged) surface~\cite{squires_bazant_2004}. 
This can play a critical role near the sharp corners found at the opening of a microfluidic channel, where the change in electric field is significant. 
Squires and Bazant~\cite{squires_bazant_2004} argued that this results in fluid vortices and localized particle aggregation. 
Such vortices were first studied by Thamida and Chang~\cite{thamida_nonlinear_2002, takhistov_chang_electrokinetic_2003} and later applied by Zehavi~\textit{et al.} to induce rapid particle accumulation and trapping~\cite{zehavi_particle_2014, zehavi_competition_2016}. 
Similar non-linear electrokinetic transport phenomena have also been reported by using isolated symmetrical sharp corners~\cite{chen_vortex_2008, eckstein_Yossifon_nonlinear_2009} or bipolar electrodes inside a straight microchannel~\cite{ wu_ac_2008, islamPerformanceImprovementAC2013, eden_modeling_2019}.

Other approaches to transporting particles through or trapping them in/near pores use linear electro-osmotic effects in combination with a secondary driving mechanism. 
For example, Cevheri and Yoda~\cite{cevheriElectrokineticallyDrivenReversible2014, yeeExperimentalObservationsBands2018} have reported that CPs were concentrated at microchannel wall and form bands when the pressure driven flow opposes electro-osmotic flow.
Lochab~\textit{et al.}~\cite{lochabDynamicsColloidalParticles2019} demonstrated that the cross-stream migration of colloids gives rise to varying spatial distributions within microchannels. They observed CP-aggregation both in fluid bulk and near the wall of the microchannel by using the inertial effect~\cite{carloInertialMicrofluidics2009} in microfluidics. 
Huang~\textit{et al.}~\cite{huang_digitally_1999} demonstrated that continuous pressure-driven flow and electro-phoresis could lead to the separation and focusing of proteins and other charged molecules in a channel.
Similarly, Stein~\textit{et al.}~\cite{stein_electrokinetic_2010} revealed that DNA-aggregation may be achieved through the interplay of electro-phoresis, electro-osmosis, and the unique statistical properties of confined polymers. 
Lastly, Rempfer~\textit{et al.} reported the use of a microcapillary pipette to concentrate $\lambda$-phage DNA at the tip of, and its subsequent delivery into, the capillary using a combination of electro-osmotic flow, pressure-driven flow, and electro-phoresis~\cite{Joost_rempfer_selective_2016, rempfer_NP_2017}. 

These linear electro-osmotic and electro-phoretic effects may be enhanced by imposing a salt gradient over the pore. 
For example, imposing a salt gradient led to a substantial improvement of the capture rate of DNA into a nanofabricated SiN pore~\cite{wanunu_electrostatic_2010, chou_enhancement_2009, hatlo_Rene_translocation_2011}. 
A salt gradient can also induce ionic diffusio-osmosis and -phoresis,~\textit{i.e.}, a tendency of charged objects to move in a salt gradient. 
Rabinowitz~\textit{et al.}~\cite{rabinowitz_nanoscale_2019} showed that similar to ICEO, this combination can lead to nanoscale fluid vortexes and non-linear electro-osmotic flow around a pipette tip. 
Recently, Shin~\textit{et al.}~\cite{PhysRevX.7.041038} used a combination of fluid flow and diffusio-phoresis to accumulate CPs at the outlet of a microchannel; the particles may be orders of magnitudes smaller than the pore size. 
These authors further showed that flow-induced particle clustering can cover the pore and lead to irreversible clogging when left unperturbed for a substantial time ($\sim 20$~minutes).

In the above examples, direct-current (DC) E-fields were predominantly used. 
However, alternating-current (AC) fields may be applied to create greater sensitivity or selectivity. 
For instance, Xuan and coworkers~\cite{zhu_electrokinetic_2012, patel_microfluidic_2012, xuan_reservoir-based_2013, patel_reservoir-based_2013, kale_joule_2014, lu_viscoelastic_2015, kale_three-dimensional_2018} demonstrated selective concentration of particles and control of their transport into a reservoir-microchannel junction using biased alternating E-fields. 
In addition, reversal of an outwardly directed E-field can be used to concentrate DNA near a pipette tip, when combined with electro-osmotic flow, electro-phoresis, and dielectro-phoretic effects~\cite{klenerman_ying_programmable_2002, ying_frequency_2004}. 
Lastly, AC fields are also suited to remove or prevent clogging. Harrison~\textit{et al.}~\cite{harrison_electrokinetic_2015} reported a high AC voltage induced-charge electro-osmosis that can concentrate particles inside channel reservoirs and avoid channel clogging. 
Such clogging may also be overcome by using well-designed geometries, like pillar arrays~\cite{wang_clog-free_2014} and diamond-shaped gradient nanopillar arrays~\cite{wang_hydrodynamics_2015, karlsen_pressure_2015}.

To summarize, there is a wealth of knowledge on the transport, concentration, and trapping of CPs in nano- and microfluidic channels and the use of geometry and AC fields to prevent clogging. 
Our work explores the combination of AC electric fields and imposed salt gradients, to contribute to the current level of understanding and we will come back to the various features, introduced here, in our discussion section. Therein, we will also argue that our numerical approach provides the essential insights to understand and qualitatively explain our experimental findings and we identify the innovative aspects of our research. 

\section*{Experimental Results}

\begin{figure}[h]
 \centering
 \includegraphics[width=\columnwidth]{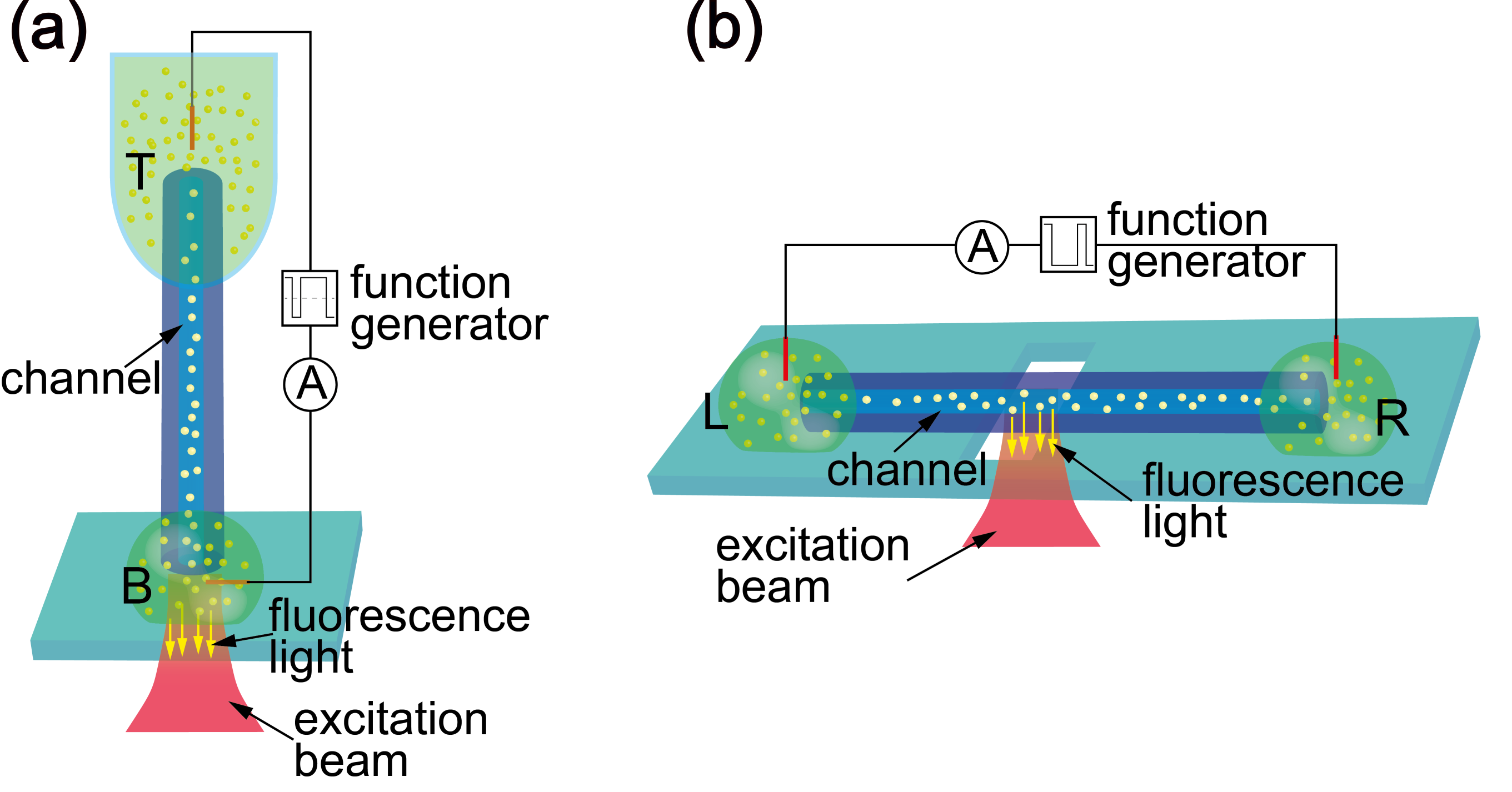}
 \caption{\label{figure:setup}Schematic representation of the two measurement configurations. (a) Vertical configuration used to observe CPs at the bottom opening of the channel; `T' and `B' labels denote top and bottom, respectively. (b) Horizontal configuration to observe the motion of the CPs inside the channel, `L' and `R' labels denote left and right, respectively. The two reservoirs (green) at either end of silica capillary (blue) contained aqueous CP (yellow) solutions and were connected to a waveform generator via immersed copper electrodes. The illuminated area is shown in red. from this region fluorescent light coming from the CPs (yellow arrows) is collected through the microscope objective.}
\end{figure}

We used two measurement configurations, vertical and horizontal, which are schematically depicted in Fig.~\ref{figure:setup}. Both feature two liquid reservoirs containing DI (de-ionized) water on one side and a sodium-chloride (NaCl) solution on the opposite side. The reservoirs are connected via a silica glass capillary ($10$~mm in length) containing the microchannel of $5~\mu$m in diameter (length to diameter ratio of $2 \times 10^3$). An electrical potential was applied at the two reservoirs using copper (Cu) electrodes, which resulted in an electrokinetic flow. We prepared for a typical measurement by filling the top (T) and left (L) reservoirs, respectively, with a solution of high salt concentration ranging from 0.25~mM to 10~mM NaCl. The respective bottom (B) and right (R) reservoirs were filled with a solution of lower salt concentration or DI water. The fluorescent CPs ($200$~nm diameter) were added to the liquid in both reservoirs before placement.

The vertical setup is shown in Fig.~\ref{figure:setup}(a). 
Here, the capillary was held vertically connecting two volumes of liquids. The T reservoir consisted of a polymer tube containing $80~\mu$L of aqueous solution. 
A droplet on a glass cover-slip acted as the B reservoir. 
We recorded the particles motion at the B opening of the microchannel using an inverted microscope in fluorescence mode. 
Fig.~\ref{figure:setup}(b) depicts the horizontal measurement configuration. Here, the glass capillary was placed horizontally on top of the microscope objective inside the immersion oil, which is necessary to obtain an aberration free image of the particles inside the channel. 
The two reservoirs at either end of the capillary (L and R, respectively) contained $40~\mu$L-droplets of the aqueous CP solution. 
The equal size of the droplets also minimizes the hydrostatic pressure difference between the two reservoirs. Because of the length of the channel, the flow caused by any residual pressure difference is orders of magnitudes slower than that caused by osmotic pressure and electro-osmotic forces.
CP traces were recorded through the microscope objective focused on the channel inside the capillary.

\begin{figure}[h]
 \centering\includegraphics[width=\columnwidth]{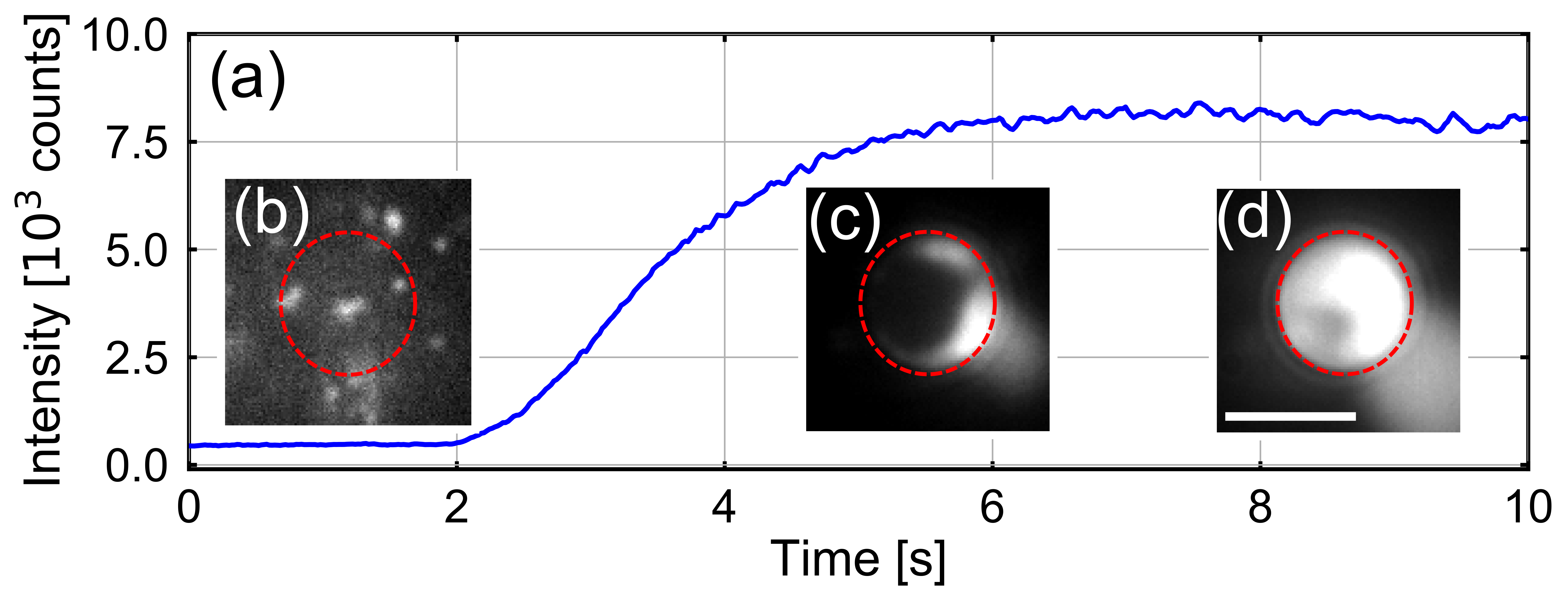}
 \caption{\label{figure:Curve}(a) Accumulated CP fluorescence intensity of (in camera counts) recorded as a function of time around the B-opening of the channel. The salt concentration in the T reservoir was $C_{\mathrm{T}} = 5$~mM, the B-reservoir was filled with DI water, and the DC applied potential across the two reservoirs was 100~V (Cathode at T and Anode at B). (b-d) Still images of the CP intensity at the B opening of the channel, indicated by the dotted circle, at different stages of aggregation. The scale bar is $5~\mu$m.}
\end{figure}

In Fig.~\ref{figure:Curve}, we present a typical measurement of the total fluorescence signal at the B opening of the vertical channel. 
Upon applying the potential, we observed the intensity of the fluorescence emission from CPs at the B opening to increase rapidly, before saturating. 
This dynamics corresponded to CP aggregation, as can be appreciated from the insets of Fig.~\ref{figure:Curve}, which contain fluorescence still images made during this process. The ESI$^{\dag}$ contains a movie that shows the associated dynamics of aggregation.

\begin{figure}[h]
 \centering\includegraphics[width=\columnwidth]{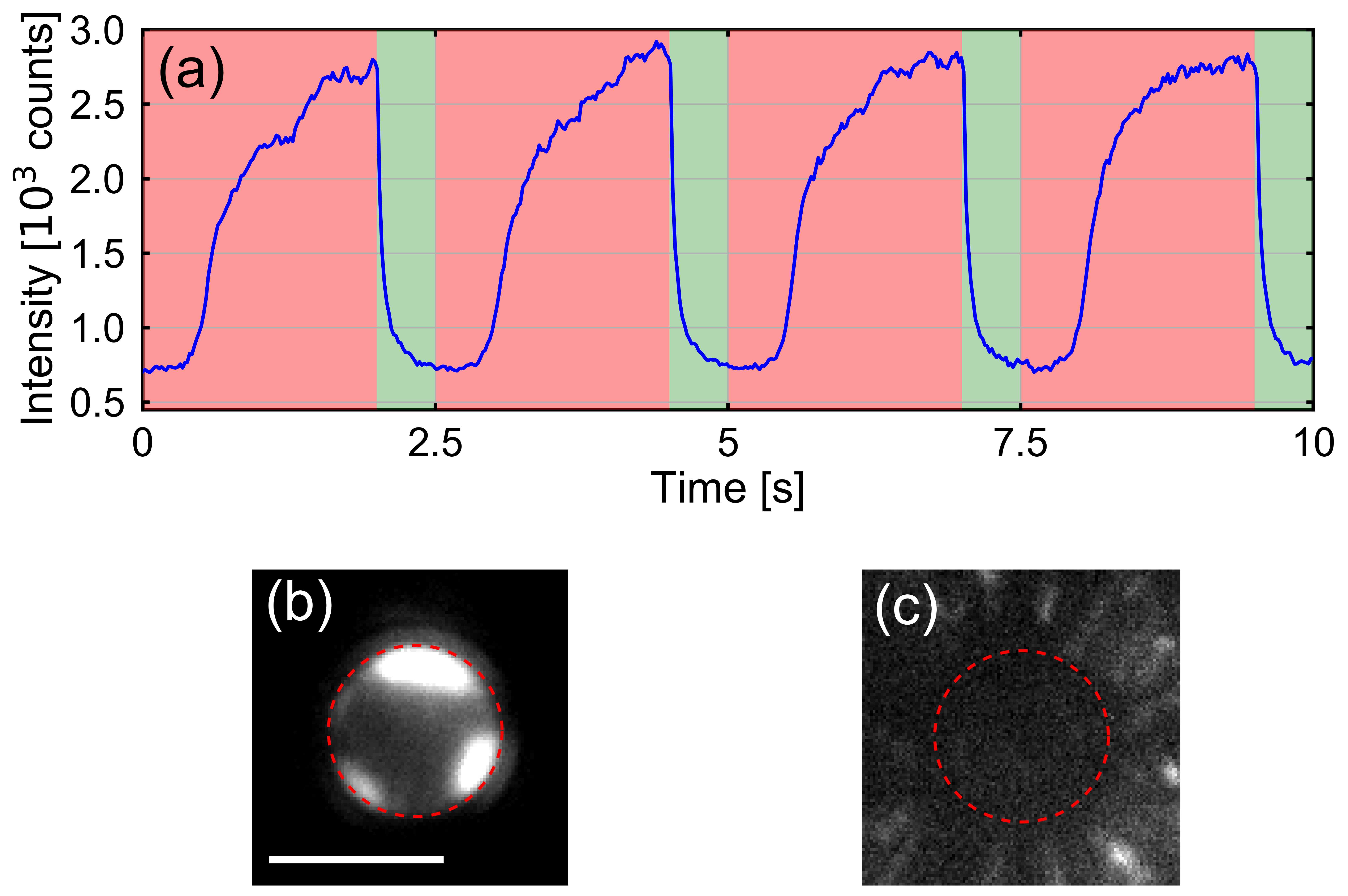}
 \caption{\label{figure:ACCurve}(a) Accumulated fluorescence intensity of CPs at the B opening while applying an asymmetric square-waveform potential, with a period of 2.5~s, duty cycle of $80\%$, and amplitude of 30~V. All other parameters are identical to those in Fig.~\ref{figure:Curve}. (b) Still image of aggregated CPs. (c) Image of the CP intensity after clearing the opening by reversing the potential. The scale bar is $5~\mu$m.}
\end{figure}

Next, we examined if the CP aggregate could be removed by inverting the applied potential.
We thus applied an asymmetric square-waveform potential to our system, with a period of 2.5~s, a duty cycle of $80\%$, and an amplitude of $30$~V. Fig.~\ref{figure:ACCurve}(a) shows the fluorescence intensity, which is an indicator of the CP accumulation, around opening B over 4 periods of this applied potential.
Fig.~\ref{figure:ACCurve}(b,c) depict representative images of the CP intensity at the B opening of the channel after clogging and clearing, respectively. The ESI$^{\dag}$ contains a movie that shows the clearance process resulting from the potential reversal.

From the fluorescence signal it is clear that the CPs started to aggregate when applying positive potential (the potential decreases from opening T to opening B) across the two reservoirs. 
The aggregation process took a few seconds, after which the fluorescence intensity remained roughly constant. 
When the potential was reversed, the CPs were quickly dispersed away from the bottom opening of the channel. 
Interestingly, the time of unclogging is, in all our observations, including a wide range of applied potentials and salt concentrations in the T reservoir, a factor of 10 smaller than the time of clogging.
The corresponding measured time-constants are plotted in Fig.~S6 of the ESI$^{\dag}$. 

This last observation proved key to keeping the channel from permanent clogging, while simultaneously obtaining a net transport of CPs, as we will describe later. 
In relation to this, it should also be noted that when the positive potential is applied for too long, we were not always able to open the pathway by reversing the EOP direction. 
Therefore, to avoid irreversible clogging, the applied potential must be reversed at an appropriate moment.

\begin{figure}
 \centering\includegraphics[width=\columnwidth]{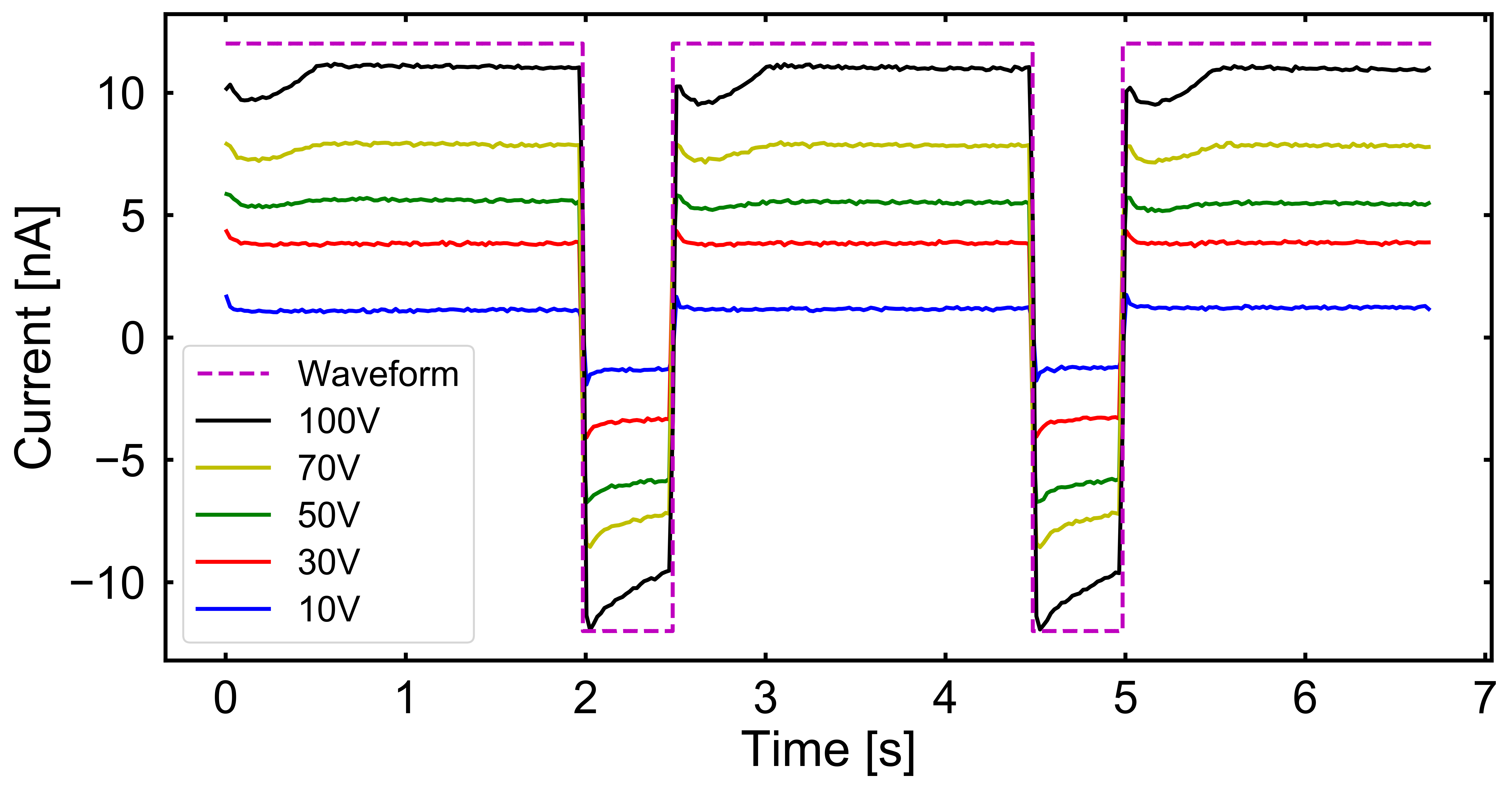}
 \caption{\label{figure:current}The measured ionic current passing through the channel when subjected to the same square-waveform potential as in Fig.~\ref{figure:ACCurve} under the same circumstances. The purple dashed line indicates the shape of the applied waveform. The current is show as a function of time for applied voltages of: 10~V (blue), 30~V (red), 50~V (green), 70~V (yellow), and 100~V (black).}
\end{figure}

We studied the impact of CP aggregation on the ionic flow through the channel by measuring the electric current passing through it. 
In Fig.~\ref{figure:current}, we plot the ionic current across the channel while applying square-waveform potentials of varying amplitude. 
Note that in all cases, there appears to be a dip in the current when the positive voltage is just applied, which saturates to a constant value. 
Upon switching to negative voltages, the current reversed and its magnitude decreased over the time interval. 
From this observation, we conclude that the clogging is predominantly affecting the CP transport and has a limited effect on the current. 

\section*{Measuring Net Particle Transport}

In the previous section, we have shown that an appropriately tuned alternating potential could be used to aggregate and disaggregate CPs at the pore opening. However, this is in itself not sufficient to show that net particle transport can be obtained. In this section, we therefore consider the trajectories of CPs inside the channel. Obtaining these trajectories necessitated a switch to our (effectively equivalent) horizontal geometry, which is shown in Fig.~\ref{figure:setup}(b). Here, the L, R openings are equivalent to the T, B openings of our vertical setup. The microscopy data is, however, taken at the mid-point of the channel, instead of the opening.

\begin{figure}[h]
 \centering\includegraphics[width=\columnwidth]{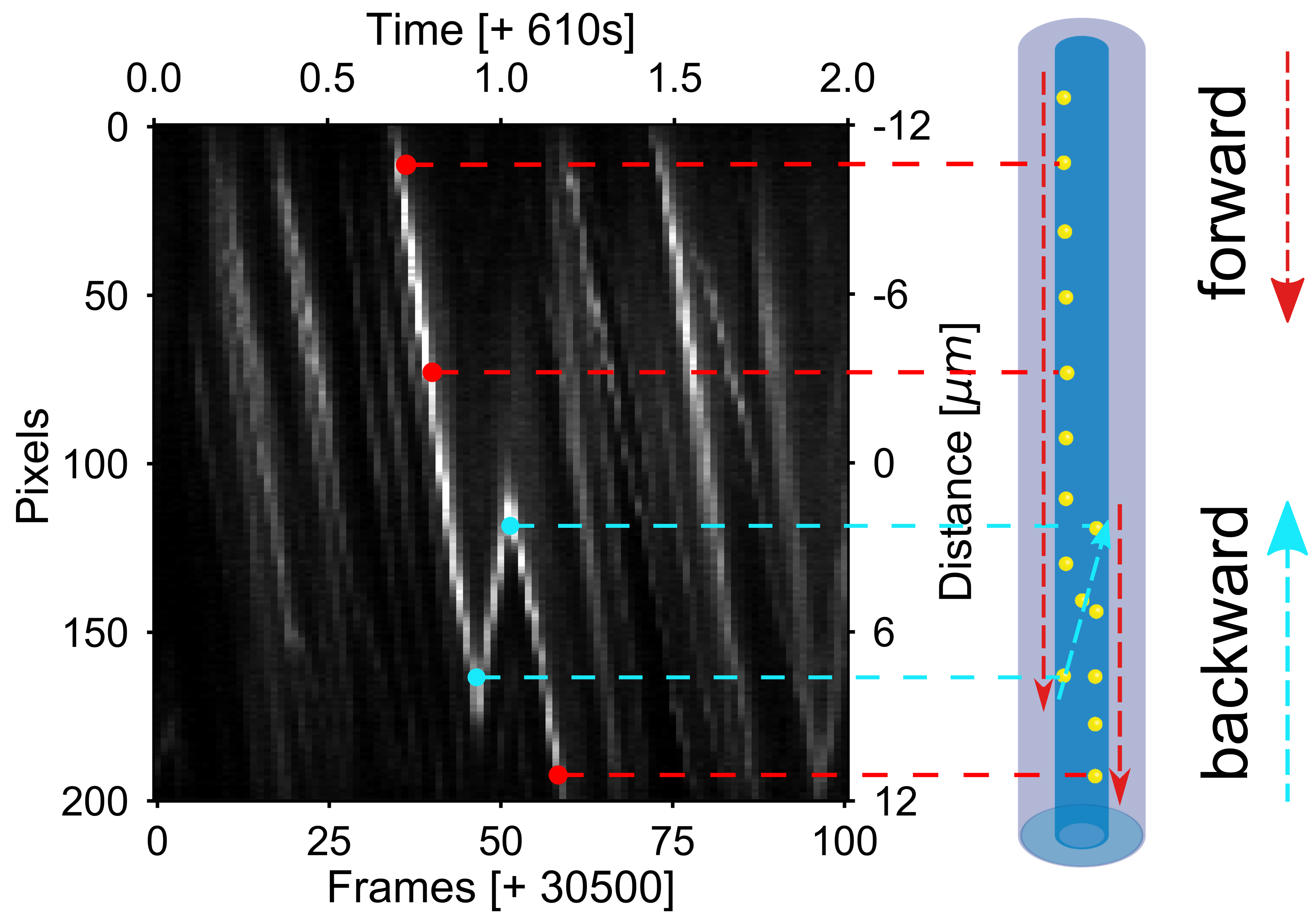}
 \caption{\label{figure:TrackerjammingAC}Kymograph showing the positions of CPs (bright due to their fluorescence) inside the channel as a function of time, referenced to the start of the measurement. The tracks were recorded after applying an asymmetric square-waveform electric potential with period of 1~s, duty cycle of $90\%$, and amplitude of $30$~V for approximately 10~minutes. The salt concentration in the L reservoir was $C_{\mathrm{L}} = 5$~mM; the R reservoir was filled with DI water. The sketch to the right of the graph illustrates how a kymograph trace can be related to a particle path.}
\end{figure}

Following the no-clogging recipe proposed in the previous section, we applied a square-waveform periodic potential with period of 1~s, duty cycle of $90\%$, and amplitude of $30$~V. The salt concentration in the L reservoir was $C_{\mathrm{L}} = 5$~mM, the R reservoir was filled with DI water. In this case, even after 10 minutes from the start of the measurement, no clogging was observed and the net motion of the CPs through the channel was preserved. 
This net flow is clearly present in the kymograph in Fig.~\ref{figure:TrackerjammingAC} that depicts particle tracks in our field of view for one period of the AC field; the ESI$^{\dag}$ contains a movie showing the corresponding tracks in real time.
Clearly, the particles can be transported through the channel. 
At the same time, we also measured a non-zero ionic current which similar to the measurements in the vertical configuration, see Fig.~S5 of the ESI$^{\dag}$.

We note in passing that by applying a DC potential for about 10 minutes, to verify clogging, we could also observe a different form of particle stagnation, akin to what has been recently reported by Shin~\textit{et al.}~\cite{PhysRevX.7.041038}. 
During this time a cluster of fluorescent CPs slowly grew into the imaging area in the middle of the channel, eventually filling the whole field of view. 
This clogging phenomenon starts from inside the channel due to a separate mechanism that is the subject of a follow-up study~\cite{zhu_under_prep}.
Once the channel had clogged in this manner, we were unable to open the pathway via EOP in the opposite direction, even at a greater applied voltage.
Therefore, to avoid clogging in presence of CPs, applying a constant potential across the channel for times exceeding a few seconds has to be avoided at any moment during the experiment.

\section*{Numerical Modeling}

We modeled our system numerically to understand the processes that cause CP clogging near the B opening.
Here, we will briefly outline the model to aid our discussion, full details are provided in our follow-up study~\cite{zhu_under_prep}. We solved the fully coupled Stokes-Poisson-Nernst-Planck (SPNP) equations for a stationary response using the COMSOL Multiphysics 5.4 suite that uses the finite-element method.
This means that for the ions: advection by the fluid, thermal diffusion, and migration in an electric field are accounted for.
From the obtained fluid velocity, salt concentration profiles, and electrostatic potential, the velocity of the CPs was computed, which incorporated advection, electro-phoresis, and ionic diffusio-phoresis, respectively.
Here, we made a point-particle approximation,~\textit{i.e.}, the presence of the particle does not significantly perturb its environment. This is a necessary reduction to keep our calculation tractable due to the wide separation of length scales involved.

\begin{figure}[!h]
 \centering\includegraphics[width=\columnwidth]{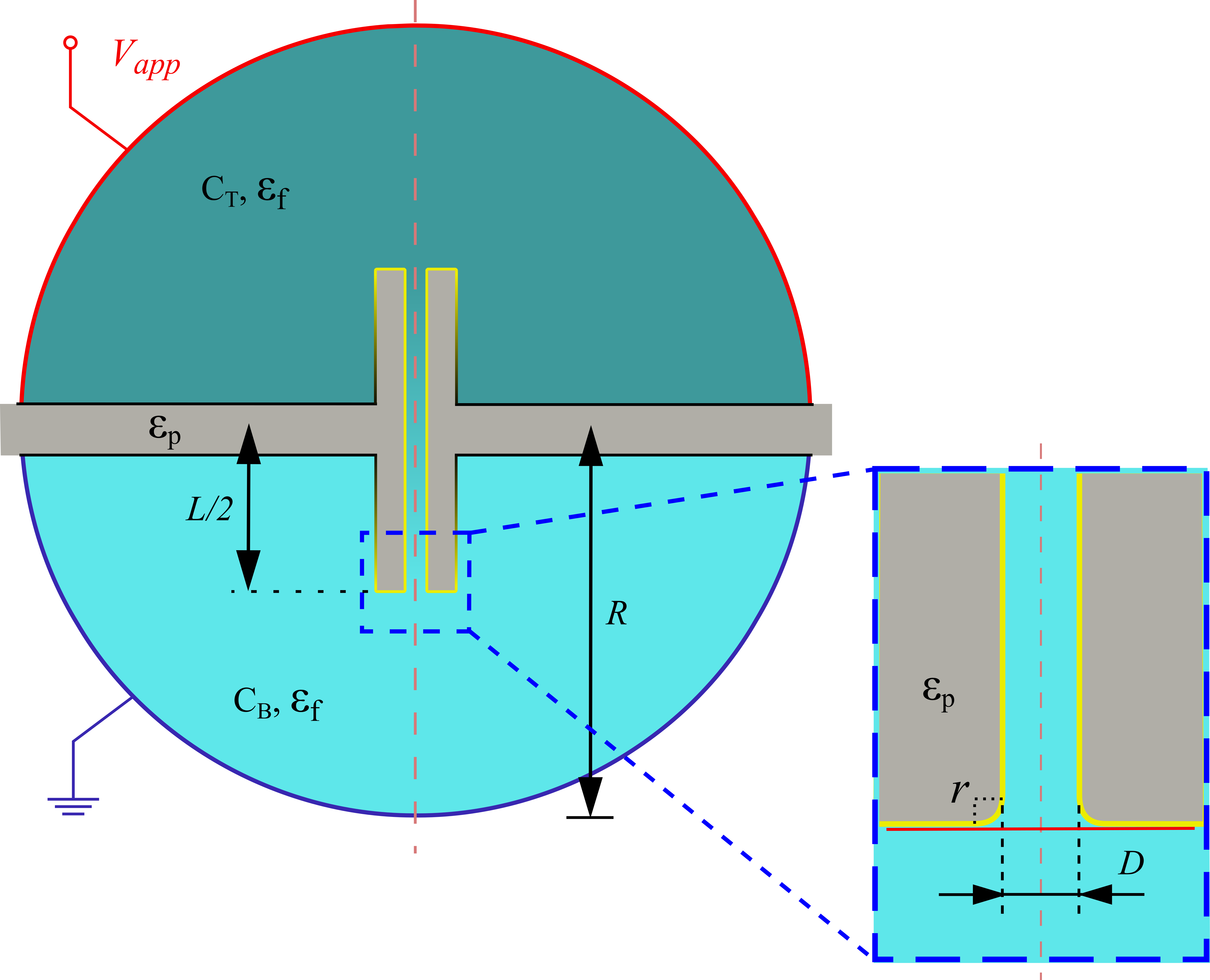}
 \caption{\label{fig:geometry}Sketch of the axisymmetric (about the central red dashed line) microchannel geometry used in our numerical modeling. The inset shows a zoom-in of a part of the geometry that is of interest. The setup consists of a top and bottom hemisphere, radius $R$, with respective salt concentrations $C_{\mathrm{T}}$ and $C_{\mathrm{B}}$ and electrostatic potential $V_{\mathrm{app}}$ (red) and ground (blue). These are separated by a divide and connected \textit{via} cylindrical pore (grey), length $L$, inner diameter $D$, and a small rounding of the edges with radius $r$. The setup is filled with suspending fluid with relative permittivity $\epsilon_{f}$ and the divide and pore are solids with relative permittivity $\epsilon_{p}$. The boundary color black indicates a no-slip, no-penetration, no-surface-charge boundary. On the boundaries marked yellow there is an osmotic surface slip velocity for the fluid.}
\end{figure} 

We use the geometry depicted in Fig.~\ref{fig:geometry} to perform these calculations with a set of parameters that best approximate the experimental setup. The sketch in Fig.~\ref{fig:geometry} is not to scale. 
Specifically, the pore has an inner diameter of $D = 2.5~\mu$m, length of $L = 10$~mm, thickness of $t = 50$~$\mu$m, the corners are rounded with a radius of $r = 500$~nm, and the divide between the reservoirs has a width of $500$~$\mu$m. The latter proved sufficient to ensure that most of the field decayed through the pore, as in the experiment. We ensured that we obtained mesh-independent results using the meshing approaches laid out in refs.~\cite{Joost_rempfer_selective_2016,rempfer_NP_2017}, also see Fig.~S7 of the ESI$^{\dag}$.

An electrostatic potential $V_{\mathrm{app}}$ and ground applied in the respective reservoirs, placed sufficiently far away from the pore to limit finite-size effects. 
On the walls, except for the inner channel walls, we use no-surface-charge boundary conditions and the fluid is treated as a dielectric medium with relative permittivity $\epsilon_{f}$. 
The divide and pore are solids with relative permittivity $\epsilon_{p}$. 
The fluid experiences a no-normal-stress condition at the reservoir edges and no-slip conditions on all the black surfaces of Fig.~\ref{fig:geometry}. 
On the yellow surfaces an effective slip velocity is imposed that captures ionic diffusio-osmosis and electro-osmosis. As a result, it is not necessary in our model to explicitly resolve the Debye screening layer. 
On the outer-edge of the pore the effective slip velocity smoothly transitions to the no-slip boundary condition on the divide, as indicated by the gradient from yellow to black. 
This improves the numerical stability of our SPNP solver.

Our system parameter choices are temperature $T = 300$~K, fluid dynamic viscosity $\eta = 10^{-3}$~Pa$\:$s and relative permittivity $\epsilon_{r} = 80$. The divide and pore have $\epsilon_{p} = 4$. For the ions we use Na$^{+}$ and Cl$^{-}$ with diffusion coefficients  $D_{+} = 1.3 \cdot 10^{-9}$~m$^{2} \:$s$^{-1}$ and $D_{-} = 2.0 \cdot 10^{-9}$~m$^{2} \:$s$^{-1}$, respectively. We varied $V_{\mathrm{app}}$ from $-100$ to $100$~V and take the zeta potential of the wall for the computation of the effective slip velocity to be $-75$~mV. 
The effective velocity of the CPs is computed using a zeta potential of $-25$~mV. 
The salt concentration in the B reservoir was fixed to $C_{\mathrm{B}} = 0.5$~mM, whilst the salt concentration in the T reservoir was set to $C_{\mathrm{T}} = 7.5$~mM.

\begin{figure}[!h]
 \centering\includegraphics[width=\columnwidth]{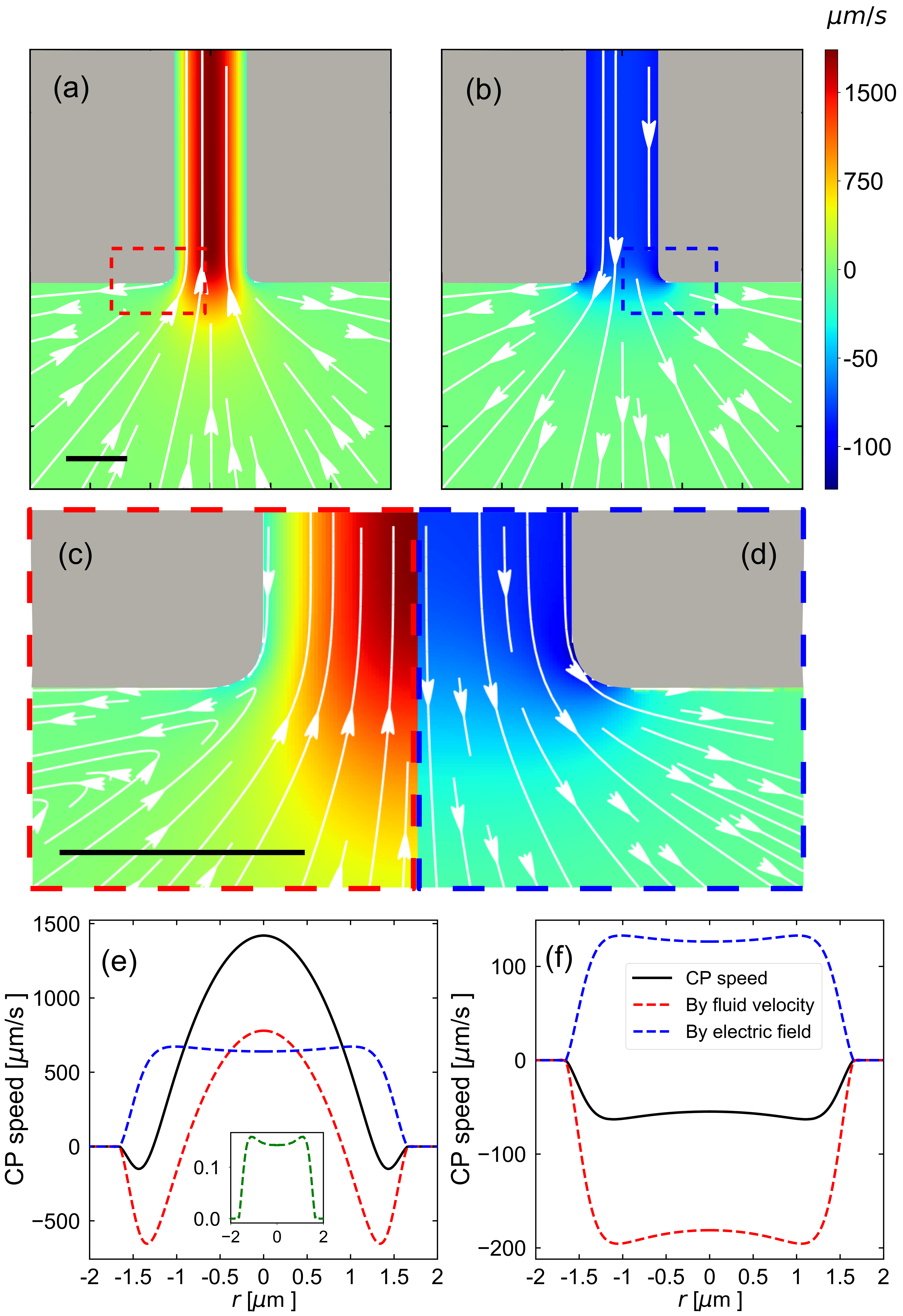}
 \caption{\label{fig:computational}Characterization of the CP velocity around the bottom pore opening. The parameters used are (a, c, e) $V_{\mathrm{app}} = 100$~V, $C_{\mathrm{T}} = 7.5$~mM,  $C_{\mathrm{B}} = 0.5$~mM; and (b, d, f) $V_{\mathrm{app}} = -100$~V, $C_{\mathrm{T}}=0.5$~mM, and $C_{\mathrm{B}} = 0.5$~mM. (a,b) The CP velocity around the bottom opening of the nanopore (gray), see Fig.~\ref{fig:geometry}. White arrows indicate the direction of the CP transport and the magnitude is indicated using the color legend. The individual figures are axisymmetric around a vertical axis running through center of the image. (c,d) A close-up view of the red and blue dashed boxes in (a,b) showing the vortices that appear in (a) and which are absent in (b). (e,f) The components to the CP speed (the sign indicates the direction along the channel) as a function of $r$ which measures the position across the bottom opening of the channel. The relevant cross section is indicated using the red line in the inset to Fig.~\ref{fig:geometry}. The solid black curves shows the total speed, the red dashed curves show the advective contribution, the blue dashed curves show the electro-phoretic component, and green curve shows the effect of ionic-diffusio-phoresis. The inset to (e) shows that the latter has a small contribution; there is no contribution in (f). All scale bars are $2~\mu$m.}
\end{figure} 

Fig.~\ref{fig:computational}(a,b) shows the CP velocity around the bottom opening of the channel. Note that the speed is predominantly directed inward close to the pore axis, when the salt concentration in the T reservoir is $C_{\mathrm{T}}=7.5$~mM. Right at the periphery of the pore opening, however, the particle flow is directed outward, see Fig.~\ref{fig:computational}(c). The particle trajectories around the pore mouth are part of a larger vortex, not shown here, a vortex that is also present in the fluid velocity field, see Fig.~S8 of the ESI$^{\dag}$. Fig.~\ref{fig:computational}b and the zoom-in in Fig.~\ref{fig:computational}d shows that when there is no salt gradient and the E-field is reversed, the vortex disappears.

Fig.~\ref{fig:computational}(e,f) show the CP speed along the bottom opening of the channel. Similar cross sections for the middle and top of the channel are provided in Fig.~S9 of the ESI$^{\dag}$. Complementary plots of the hydrostatic pressure along the pore axis and around the B opening are provided in Fig.~S10 of the  ESI$^{\dag}$. Turning back to Fig.~\ref{fig:computational}, there is a clear competition between advection of the CPs by the fluid and their electro-phoresis in E-field through the pore. This results in a reversal of the direction of CP motion near the wall compared to the center line, which is the origin of the observed vortexes around the pore. These plots further show that the salt gradient has almost no direct influence on the motion of the CPs. 
However, as can be appreciated by contrasting this result with Fig.~\ref{fig:computational}(f), the presence of a salt gradient strongly modulates the way the E-field is shaped and decays throughout the pore, thereby having a dominant, albeit indirect effect. N.B. If only the E-field had been flipped in Fig.~\ref{fig:computational}(f), graph (e) would be speed-wise inverted.

\section*{Discussion}

The combined experimental and numerical results lead us to conclude that CP aggregation and clogging of the pore can occur via a combination of advection, driven by electro-osmotic flow, and electro-phoresis. Both of these effects are strongly modulated by the presence of a salt gradient, which was found to be a necessary ingredient to the clogging of the pore opening. The second condition is that the applied potential is upstream with the salt concentration gradient. In this case, clogging of the channel occurs at the low-salt-concentration side.

The clogging of the opening of the channel was further found to only strongly affect the transport of CPs, as no appreciable change of ionic current was observed after CP aggregation. To explain this observation, we estimate the Dukhin length based on the definition $l_{Du} = \sigma/(e C)$, with $\sigma$ the surface charge density, $e$ the elementary charge, and $C$ the bulk (monovalent) ion concentration~\cite{bocquet2010nanofluidics}.
Assuming a salt concentration of 0.1~mM for the DI water (an upper estimate) and a surface charge density of $\sigma = 0.3 e /\mathrm{nm}^2$ (typical number for Silica glass), the Dukhin length at the B reservoir is roughly 5~$\mu$m and hence comparable with the channel diameter. This estimation indicates that the surface conductance plays a major role in the measured ionic current, while the CPs mainly influence the bulk conductivity.

The slight change in electrical conductivity of the channel under negative bias (see Fig.~\ref{figure:current}) is followed by an equally long drop in the current signal after reversing the bias.
This non-stationary behavior can be explained by considering the plug flow of liquid from the low concentration reservoir into the opening B of the channel, which is subsequently pushed out upon reversing the bias. 
The lower salt concentration in the B reservoir temporarily decreases the electric conductance of the channel, hence the decrease in the measured current. 
The duration of this negative bias is, however, not long enough for reaching an equilibrium with the higher salt concentration inside the channel. 
Therefore, under positive bias, the same portion of low-salt concentration liquid is pushed out of the channel, causing a short dip in conductance at beginning of each cycle. This is the reason for providing results with different `effective' salt gradients in our Numerical Modeling section, as will become clear shortly.

Our finite-element based calculations revealed that under the application of a salt gradient, vortex like motion of the particles occurs close to the opening. This effect is reminiscent of the results by Rabinowitz~\textit{et al.}~\cite{rabinowitz_nanoscale_2019}, as well as the type of motion that emerges due to ICEO~\cite{squires_bazant_2004}. It suggests that the initial CP aggregation at the pore opening may be attributed to this vortex-like motion, as depicted in Fig.~\ref{fig:computational}(e). A notion that is supported by a lack of aggregation when there is no salt gradient in the experiment, combined with the absence of a balance in the two driving components in the numerical calculations. Furthermore, reversing the bias temporarily reduces the salt gradient at the opening of the channel by pulling in the low-salinity solution. As a result, the vortex-like driving force on the CPs is replaced by another flow pattern that contain no nodes, similar to the pattern expected from stationary results for no salt gradient, which is depicted in Fig.~\ref{fig:computational}(f).
This outward flow can result in clearing all the CPs aggregated at the opening.

The main outcome of our investigation is that the flow pattern at the opening of the channel strongly depends on the salt-gradient, as identified by our experiments and verified by numerical modelling. 
Since reaching the stationary state can take several minutes, the exact calculation of influx and outflux of CPs will require time-dependent modeling of the flow pattern. However, the separation of length scales relevant to our problem proved prohibitive, in spite of the significant computational power at our disposal. Such modeling will therefore be left for future work.
Nevertheless, from these stationary calculations we can qualitatively understand the competition between different transport processes on CPs, which can direct us in choosing proper experimental parameters. 

\section*{Summary and Outlook}

Summarizing, we have experimentally and numerically studied the electro-osmotic aggregation of nanoparticles at the opening of a micrometer-sized silica channel in the presence of a salt gradient. Under the application of a DC electric field, aggregation of CPs at the pore opening occurred, which led to clogging of the channel. Based on numerical modelling, we explain the formation of the aggregate in terms of a balance between advective and phoretic forces. The clogging may be undone for sufficiently small aggregates through reversal of the potential. This insight allowed us to formulate a recipe for continuous nanoparticle transport by applying a square-waveform electric potential with an appropriately tuned duty cycle. Such a practical recipe is relevant to applications requiring continuous EOP,~\textit{e.g.}, sensing using membrane nanopores. 

A full understanding of the time-dependence of this process is still lacking and will be the focus of future efforts. 
However, considering the wide use of EOP in micro/nanofluidic devices, we are confident that our present results will already provide a new handle on the design and operation of such devices. The description we have provided for the underlying physical processes can further be useful for engineering situations where controlled aggregation can be advantageous, for example, for concentrating proteins in a well-defined location or for inducing crystallization by the long-period aggregation of (nano)particles. 

\section*{Methods}

\subsection*{Preparation of the Channel and Fluorescent Colloids}

\textbf{Silica capillary (microchannel):} The Silica glass capillary (Fused Silica Capillary, CM Scientific, Silsden, UK) used in our experiments had an inner-channel diameter of 5~$\mu$m and an outer diameter of 150~$\mu m$. It was coated with a 12~$\mu$m thick protective layer on its outer surface. In order to receive the fluorescent light from our CPs inside the channel, we removed the opaque protective layer and wiped the capillary surface using a tissue wet with ethanol and isopropanol before constructing the sample cell. Finally, we cut the capillary to 10~mm using a diamond cutter.

\textbf{Preparation of the sample:} \textit{The vertical configuration} We put the prepared capillary into the top reservoir which had a 0.5~mm diameter hole in it. We subsequently added optical epoxy (NOA61, Thorlabs) in the gap between the capillary and the hole, and used a UV-light source (CS2010, Thorlabs) cure the glue, thereby fixing the capillary to the top reservoir. \textit{The horizontal configuration} We put the prepared capillary in the middle of the plastic microscope slide which contains a rectangular opening of $1\times2~\mathrm{mm}^{2}$ area. We used the optical epoxy (Griffon combi fast, Bolton Adhesives) between capillary and slide at the edge of the hole to fixed the capillary and to prevent the solution in the reservoirs from mixing with the immersion oil on top of the objective.

\textbf{Fluorescent Colloids:} The fluorescent CPs (sicastar-greenF, micromod Partikeltechnologie GmbH) were produced by hydrolysis of orthosilicates and related compounds. They have a hydrophilic surface with terminal Si-OH-groups. The CPs were monodisperse with a mean diameter of 200~nm and nonporous with with a density of $50$~mg/ml. They emit florescent light with wavelength of $510$~nm when they are excited by a laser with a wavelength of $485$~nm. The concentration of the CPs was controlled by diluting the stock solution with DI water. In our experiments, the CPs were suspended in different salt solutions, prepared by diluting a stock solution of $100$~mM NaCl ($\geq 99.5\%$, Sigma-Aldrich) to their final salt concentrations.

\subsection*{Microscopy Measurement Setup}

Light from a solid-state single-mode laser (Coherent OBIS, $488$~nm) was used to illuminate the observation area. In the vertical setup, we observed the CPs around the bottom opening, at a distance of roughly $5~\mu$m from the end of the channel. This enabled us to study the channel cross section. In the horizontal setup, we instead covered in the middle of the channel in a direction along the axis, allowing us to see CP traces. In both cases, we used a home-built microscope with a 100~$\times$ 1.30 NA Olympus oil-immersion objective, 100~$\times$ overall magnification, and an effective field of view of around $200~\times 200~\mu$m. The fluorescent light from the diffusing particles was recorded using a sCMOS camera (Hamamatsu Orca Flash4).

\subsection*{Measurement preparation procedures}

A small drop of solution without CPs was placed in the top reservoir for the vertical configuration setup or on the one end of the capillary for the horizontal setup. We then waited for 2 to 3 minutes until the liquid was pulled into the whole channel by capillary forces. After this, another drop of solution was placed on the other end of the capillary. An electrical potential with a maximum amplitude of 100~V was applied to our setup using Cu electrodes connected in series to a commercial current amplifier (DLPCA-200, FEMTO), an oscilloscope, and a data acquisition card (NI 6216-USB, National Instruments). The combination of these allowed us to simultaneously view and record ionic currents. By monitoring the ionic currents, we could confirm that the two reservoirs at either side of the channel were connected through the capillary. After establishing the electric connection, the solution in the top and bottom reservoir was replaced with drops of suspension containing the CPs. Square-waveform potentials in the range $-100$~V to $+100$~V with a range of duty cycles were applied across the channel, whilst the particle motion was recorded through the microscope.

\section*{Conflicts of interest}

There are no conflicts to declare.

\section*{Acknowledgements}
This research was supported by the Netherlands Organization for Scientific Research (NWO grant 680.91.16.03). J.d.G. thanks NWO for funding through StartUp Grant 740.018.013. J.d.G. further acknowledges financial support through association with the EU-FET project NANOPHLOW (766972) within Horizon 2020. Z.Z. thanks the China Scholarship Council (CSC) for financial support through Grant No.201806890015. The authors thank Paul Jurrius and Dante Killian for technical support and Ben Werkhoven for fruitful discussions.



\balance


\bibliography{Bib} 

\providecommand*{\mcitethebibliography}{\thebibliography}
\csname @ifundefined\endcsname{endmcitethebibliography}
{\let\endmcitethebibliography\endthebibliography}{}
\begin{mcitethebibliography}{63}
\providecommand*{\natexlab}[1]{#1}
\providecommand*{\mciteSetBstSublistMode}[1]{}
\providecommand*{\mciteSetBstMaxWidthForm}[2]{}
\providecommand*{\mciteBstWouldAddEndPuncttrue}
  {\def\EndOfBibitem{\unskip.}}
\providecommand*{\mciteBstWouldAddEndPunctfalse}
  {\let\EndOfBibitem\relax}
\providecommand*{\mciteSetBstMidEndSepPunct}[3]{}
\providecommand*{\mciteSetBstSublistLabelBeginEnd}[3]{}
\providecommand*{\EndOfBibitem}{}
\mciteSetBstSublistMode{f}
\mciteSetBstMaxWidthForm{subitem}
{(\emph{\alph{mcitesubitemcount}})}
\mciteSetBstSublistLabelBeginEnd{\mcitemaxwidthsubitemform\space}
{\relax}{\relax}

\bibitem[Chowdhury \emph{et~al.}(2019)Chowdhury, Salam, Debnath, Islam, and
  Saha]{CHOWDHURY2019265}
A.~H. Chowdhury, N.~Salam, R.~Debnath, S.~M. Islam and T.~Saha, in
  \emph{Nanomaterials Synthesis}, ed. Y.~B. Pottathara], S.~Thomas,
  N.~Kalarikkal, Y.~Grohens and V.~Kokol, Elsevier, 2019, pp. 265--294\relax
\mciteBstWouldAddEndPuncttrue
\mciteSetBstMidEndSepPunct{\mcitedefaultmidpunct}
{\mcitedefaultendpunct}{\mcitedefaultseppunct}\relax
\EndOfBibitem
\bibitem[Driscoll and Delmotte(2019)]{driscollLeveragingCollectiveEffects2019}
M.~Driscoll and B.~Delmotte, \emph{Curr Opin Colloid Interface Sci}, 2019,
  \textbf{40}, 42--57\relax
\mciteBstWouldAddEndPuncttrue
\mciteSetBstMidEndSepPunct{\mcitedefaultmidpunct}
{\mcitedefaultendpunct}{\mcitedefaultseppunct}\relax
\EndOfBibitem
\bibitem[Akbari \emph{et~al.}(2016)Akbari, Rahim, Ehrlicher, and
  Ghoshal]{akbariGrowthAttachmentFacilitatedEntry2016}
A.~Akbari, A.~A. Rahim, A.~J. Ehrlicher and S.~Ghoshal, \emph{Environ. Sci.
  Technol. Lett.}, 2016, \textbf{3}, 399--403\relax
\mciteBstWouldAddEndPuncttrue
\mciteSetBstMidEndSepPunct{\mcitedefaultmidpunct}
{\mcitedefaultendpunct}{\mcitedefaultseppunct}\relax
\EndOfBibitem
\bibitem[Nath \emph{et~al.}(2018)Nath, Dave, and
  Patel]{nathRevisitingRecentApplications2018}
K.~Nath, H.~K. Dave and T.~M. Patel, \emph{Trends Food Sci. Technol.}, 2018,
  \textbf{73}, 12--24\relax
\mciteBstWouldAddEndPuncttrue
\mciteSetBstMidEndSepPunct{\mcitedefaultmidpunct}
{\mcitedefaultendpunct}{\mcitedefaultseppunct}\relax
\EndOfBibitem
\bibitem[Sparreboom \emph{et~al.}(2009)Sparreboom, {van den Berg}, and
  Eijkel]{sparreboomPrinciplesApplicationsNanofluidic2009}
W.~Sparreboom, A.~{van den Berg} and J.~C.~T. Eijkel, \emph{Nat. Nanotechnol.},
  2009, \textbf{4}, 713--720\relax
\mciteBstWouldAddEndPuncttrue
\mciteSetBstMidEndSepPunct{\mcitedefaultmidpunct}
{\mcitedefaultendpunct}{\mcitedefaultseppunct}\relax
\EndOfBibitem
\bibitem[Wang \emph{et~al.}(2009)Wang, Cheng, Wang, and
  Liu]{wang_electroosmotic_2009}
X.~Wang, C.~Cheng, S.~Wang and S.~Liu, \emph{Microfluid Nanofluidics}, 2009,
  \textbf{6}, 145--145\relax
\mciteBstWouldAddEndPuncttrue
\mciteSetBstMidEndSepPunct{\mcitedefaultmidpunct}
{\mcitedefaultendpunct}{\mcitedefaultseppunct}\relax
\EndOfBibitem
\bibitem[Peng and Li(2016)]{peng_electroosmotic_2016}
R.~Peng and D.~Li, \emph{Nanoscale}, 2016, \textbf{8}, 12237--12246\relax
\mciteBstWouldAddEndPuncttrue
\mciteSetBstMidEndSepPunct{\mcitedefaultmidpunct}
{\mcitedefaultendpunct}{\mcitedefaultseppunct}\relax
\EndOfBibitem
\bibitem[Han and Park(2016)]{han_optoelectrofluidic_2016}
D.~Han and J.-K. Park, \emph{Lab Chip}, 2016, \textbf{16}, 1189--1196\relax
\mciteBstWouldAddEndPuncttrue
\mciteSetBstMidEndSepPunct{\mcitedefaultmidpunct}
{\mcitedefaultendpunct}{\mcitedefaultseppunct}\relax
\EndOfBibitem
\bibitem[Li \emph{et~al.}(2019)Li, Wang, Pu, and Liu]{LI20191}
L.~Li, X.~Wang, Q.~Pu and S.~Liu, \emph{Anal. Chim. Acta}, 2019, \textbf{1060},
  1 -- 16\relax
\mciteBstWouldAddEndPuncttrue
\mciteSetBstMidEndSepPunct{\mcitedefaultmidpunct}
{\mcitedefaultendpunct}{\mcitedefaultseppunct}\relax
\EndOfBibitem
\bibitem[Ramos \emph{et~al.}(1998)Ramos, Morgan, Green, and
  Castellanos]{ramosAcElectrokineticsReview1998}
A.~Ramos, H.~Morgan, N.~G. Green and A.~Castellanos, \emph{J. Phys. D: Appl.
  Phys.}, 1998, \textbf{31}, 2338--2353\relax
\mciteBstWouldAddEndPuncttrue
\mciteSetBstMidEndSepPunct{\mcitedefaultmidpunct}
{\mcitedefaultendpunct}{\mcitedefaultseppunct}\relax
\EndOfBibitem
\bibitem[Hunter(2001)]{hunterFoundationsColloidScience2001}
R.~J. Hunter, \emph{Foundations of Colloid Science}, Oxford University Press,
  2001\relax
\mciteBstWouldAddEndPuncttrue
\mciteSetBstMidEndSepPunct{\mcitedefaultmidpunct}
{\mcitedefaultendpunct}{\mcitedefaultseppunct}\relax
\EndOfBibitem
\bibitem[Tallarek \emph{et~al.}(2000)Tallarek, Rapp, Scheenen, Bayer, and
  Van~As]{Tallarek_2000}
U.~Tallarek, E.~Rapp, T.~Scheenen, E.~Bayer and H.~Van~As, \emph{Anal. Chem.},
  2000, \textbf{72}, 2292--2301\relax
\mciteBstWouldAddEndPuncttrue
\mciteSetBstMidEndSepPunct{\mcitedefaultmidpunct}
{\mcitedefaultendpunct}{\mcitedefaultseppunct}\relax
\EndOfBibitem
\bibitem[Yuan and Oleschuk(2018)]{Xiong_yuan_Anvanced_2017}
X.~Yuan and R.~D. Oleschuk, \emph{Anal. Chem.}, 2018, \textbf{90},
  283--301\relax
\mciteBstWouldAddEndPuncttrue
\mciteSetBstMidEndSepPunct{\mcitedefaultmidpunct}
{\mcitedefaultendpunct}{\mcitedefaultseppunct}\relax
\EndOfBibitem
\bibitem[Chen \emph{et~al.}(2017)Chen, Lynch, Ren, Jia, Yang, Lu, and
  Liu]{Apeng_Femto_liter_2017}
A.~Chen, K.~B. Lynch, J.~Ren, Z.~Jia, Y.~Yang, J.~J. Lu and S.~Liu, \emph{Anal.
  Chem.}, 2017, \textbf{89}, 10806--10812\relax
\mciteBstWouldAddEndPuncttrue
\mciteSetBstMidEndSepPunct{\mcitedefaultmidpunct}
{\mcitedefaultendpunct}{\mcitedefaultseppunct}\relax
\EndOfBibitem
\bibitem[Kwon and Kim(2013)]{Kilsung_micropower_2012}
K.~Kwon and D.~Kim, \emph{J. Power Sources}, 2013, \textbf{221}, 172 --
  176\relax
\mciteBstWouldAddEndPuncttrue
\mciteSetBstMidEndSepPunct{\mcitedefaultmidpunct}
{\mcitedefaultendpunct}{\mcitedefaultseppunct}\relax
\EndOfBibitem
\bibitem[Yang \emph{et~al.}(2019)Yang, Su, Wang, and
  Wu]{yangLowvoltageEfficientElectroosmotic2019}
Q.~Yang, B.~Su, Y.~Wang and W.~Wu, \emph{Electrophoresis}, 2019, \textbf{40},
  2149--2156\relax
\mciteBstWouldAddEndPuncttrue
\mciteSetBstMidEndSepPunct{\mcitedefaultmidpunct}
{\mcitedefaultendpunct}{\mcitedefaultseppunct}\relax
\EndOfBibitem
\bibitem[Yang \emph{et~al.}(2001)Yang, Fu, and
  Lin]{yangElectroosmoticFlowMicrochannels2001}
R.-J. Yang, L.-M. Fu and Y.-C. Lin, \emph{J. Colloid Interface Sci.}, 2001,
  \textbf{239}, 98--105\relax
\mciteBstWouldAddEndPuncttrue
\mciteSetBstMidEndSepPunct{\mcitedefaultmidpunct}
{\mcitedefaultendpunct}{\mcitedefaultseppunct}\relax
\EndOfBibitem
\bibitem[Cummings and Singh(2003)]{Cummings_Diele_2003}
E.~B. Cummings and A.~K. Singh, \emph{Anal. Chem.}, 2003, \textbf{75},
  4724--4731\relax
\mciteBstWouldAddEndPuncttrue
\mciteSetBstMidEndSepPunct{\mcitedefaultmidpunct}
{\mcitedefaultendpunct}{\mcitedefaultseppunct}\relax
\EndOfBibitem
\bibitem[Lapizco-Encinas \emph{et~al.}(2004)Lapizco-Encinas, Simmons, Cummings,
  and Fintschenko]{Lapizco-Encinas_dielec_2004}
B.~H. Lapizco-Encinas, B.~A. Simmons, E.~B. Cummings and Y.~Fintschenko,
  \emph{Anal. Chem.}, 2004, \textbf{76}, 1571--1579\relax
\mciteBstWouldAddEndPuncttrue
\mciteSetBstMidEndSepPunct{\mcitedefaultmidpunct}
{\mcitedefaultendpunct}{\mcitedefaultseppunct}\relax
\EndOfBibitem
\bibitem[Barrett \emph{et~al.}(2005)Barrett, Skulan, Singh, Cummings, and
  Fiechtner]{Barrett_Dielec_2005}
L.~M. Barrett, A.~J. Skulan, A.~K. Singh, E.~B. Cummings and G.~J. Fiechtner,
  \emph{Anal. Chem.}, 2005, \textbf{77}, 6798--6804\relax
\mciteBstWouldAddEndPuncttrue
\mciteSetBstMidEndSepPunct{\mcitedefaultmidpunct}
{\mcitedefaultendpunct}{\mcitedefaultseppunct}\relax
\EndOfBibitem
\bibitem[Baylon-Cardiel \emph{et~al.}(2010)Baylon-Cardiel, Jesús-Pérez,
  Chávez-Santoscoy, and Lapizco-Encinas]{Baylon-Cardiel_Diele_2010}
J.~L. Baylon-Cardiel, N.~M. Jesús-Pérez, A.~V. Chávez-Santoscoy and B.~H.
  Lapizco-Encinas, \emph{Lab Chip}, 2010, \textbf{10}, 3235--3242\relax
\mciteBstWouldAddEndPuncttrue
\mciteSetBstMidEndSepPunct{\mcitedefaultmidpunct}
{\mcitedefaultendpunct}{\mcitedefaultseppunct}\relax
\EndOfBibitem
\bibitem[Zhu \emph{et~al.}(2009)Zhu, Tzeng, Hu, and Xuan]{zhu_dc_2009}
J.~Zhu, T.-R.~J. Tzeng, G.~Hu and X.~Xuan, \emph{Microfluid Nanofluidics},
  2009, \textbf{7}, 751\relax
\mciteBstWouldAddEndPuncttrue
\mciteSetBstMidEndSepPunct{\mcitedefaultmidpunct}
{\mcitedefaultendpunct}{\mcitedefaultseppunct}\relax
\EndOfBibitem
\bibitem[Zhu \emph{et~al.}(2010)Zhu, Tzeng, and Xuan]{zhu_continuous_2010}
J.~Zhu, T.-R.~J. Tzeng and X.~Xuan, \emph{{Electrophoresis}}, 2010,
  \textbf{31}, 1382--1388\relax
\mciteBstWouldAddEndPuncttrue
\mciteSetBstMidEndSepPunct{\mcitedefaultmidpunct}
{\mcitedefaultendpunct}{\mcitedefaultseppunct}\relax
\EndOfBibitem
\bibitem[Zhu and Xuan(2011)]{zhu_curvature-induced_2011}
J.~Zhu and X.~Xuan, \emph{Biomicrofluidics}, 2011, \textbf{5}, 024111\relax
\mciteBstWouldAddEndPuncttrue
\mciteSetBstMidEndSepPunct{\mcitedefaultmidpunct}
{\mcitedefaultendpunct}{\mcitedefaultseppunct}\relax
\EndOfBibitem
\bibitem[Zhu \emph{et~al.}(2011)Zhu, Canter, Keten, Vedantam, Tzeng, and
  Xuan]{zhu_continuous-flow_2011}
J.~Zhu, R.~C. Canter, G.~Keten, P.~Vedantam, T.-R.~J. Tzeng and X.~Xuan,
  \emph{Microfluid Nanofluidics}, 2011, \textbf{11}, 743--752\relax
\mciteBstWouldAddEndPuncttrue
\mciteSetBstMidEndSepPunct{\mcitedefaultmidpunct}
{\mcitedefaultendpunct}{\mcitedefaultseppunct}\relax
\EndOfBibitem
\bibitem[Squires and Bazant(2004)]{squires_bazant_2004}
T.~M. Squires and M.~Z. Bazant, \emph{J. Fluid Mech.}, 2004, \textbf{509},
  217–252\relax
\mciteBstWouldAddEndPuncttrue
\mciteSetBstMidEndSepPunct{\mcitedefaultmidpunct}
{\mcitedefaultendpunct}{\mcitedefaultseppunct}\relax
\EndOfBibitem
\bibitem[Thamida and Chang(2002)]{thamida_nonlinear_2002}
S.~K. Thamida and H.-C. Chang, \emph{Phys. Fluids}, 2002, \textbf{14},
  4315--4328\relax
\mciteBstWouldAddEndPuncttrue
\mciteSetBstMidEndSepPunct{\mcitedefaultmidpunct}
{\mcitedefaultendpunct}{\mcitedefaultseppunct}\relax
\EndOfBibitem
\bibitem[Takhistov \emph{et~al.}(2003)Takhistov, Duginova, and
  Chang]{takhistov_chang_electrokinetic_2003}
P.~Takhistov, K.~Duginova and H.-C. Chang, \emph{J. Colloid Interface Sci.},
  2003, \textbf{263}, 133--143\relax
\mciteBstWouldAddEndPuncttrue
\mciteSetBstMidEndSepPunct{\mcitedefaultmidpunct}
{\mcitedefaultendpunct}{\mcitedefaultseppunct}\relax
\EndOfBibitem
\bibitem[Zehavi and Yossifon(2014)]{zehavi_particle_2014}
M.~Zehavi and G.~Yossifon, \emph{Phys. Fluids}, 2014, \textbf{26}, 082002\relax
\mciteBstWouldAddEndPuncttrue
\mciteSetBstMidEndSepPunct{\mcitedefaultmidpunct}
{\mcitedefaultendpunct}{\mcitedefaultseppunct}\relax
\EndOfBibitem
\bibitem[Zehavi \emph{et~al.}(2016)Zehavi, Boymelgreen, and
  Yossifon]{zehavi_competition_2016}
M.~Zehavi, A.~Boymelgreen and G.~Yossifon, \emph{Phys. Rev. Applied}, 2016,
  \textbf{5}, 044013\relax
\mciteBstWouldAddEndPuncttrue
\mciteSetBstMidEndSepPunct{\mcitedefaultmidpunct}
{\mcitedefaultendpunct}{\mcitedefaultseppunct}\relax
\EndOfBibitem
\bibitem[Chen and Yang(2008)]{chen_vortex_2008}
J.~K. Chen and R.-J. Yang, \emph{Microfluid Nanofluidics}, 2008, \textbf{5},
  719--725\relax
\mciteBstWouldAddEndPuncttrue
\mciteSetBstMidEndSepPunct{\mcitedefaultmidpunct}
{\mcitedefaultendpunct}{\mcitedefaultseppunct}\relax
\EndOfBibitem
\bibitem[Eckstein \emph{et~al.}(2009)Eckstein, Yossifon, Seifert, and
  Miloh]{eckstein_Yossifon_nonlinear_2009}
Y.~Eckstein, G.~Yossifon, A.~Seifert and T.~Miloh, \emph{J. Colloid Interface
  Sci.}, 2009, \textbf{338}, 243--249\relax
\mciteBstWouldAddEndPuncttrue
\mciteSetBstMidEndSepPunct{\mcitedefaultmidpunct}
{\mcitedefaultendpunct}{\mcitedefaultseppunct}\relax
\EndOfBibitem
\bibitem[Wu(2018)]{wu_ac_2008}
J.~J. Wu, \emph{J. Appl. Phys.}, 2018, \textbf{103}, 024907\relax
\mciteBstWouldAddEndPuncttrue
\mciteSetBstMidEndSepPunct{\mcitedefaultmidpunct}
{\mcitedefaultendpunct}{\mcitedefaultseppunct}\relax
\EndOfBibitem
\bibitem[Islam and Askari(2013)]{islamPerformanceImprovementAC2013}
N.~Islam and D.~Askari, \emph{Microfluid Nanofluid}, 2013, \textbf{14},
  627--635\relax
\mciteBstWouldAddEndPuncttrue
\mciteSetBstMidEndSepPunct{\mcitedefaultmidpunct}
{\mcitedefaultendpunct}{\mcitedefaultseppunct}\relax
\EndOfBibitem
\bibitem[Eden \emph{et~al.}(2019)Eden, Scida, Arroyo-Currás, Eijkel, Meinhart,
  and Pennathur]{eden_modeling_2019}
A.~Eden, K.~Scida, N.~Arroyo-Currás, J.~C.~T. Eijkel, C.~D. Meinhart and
  S.~Pennathur, \emph{J. Phys. Chem. C}, 2019, \textbf{123}, 5353--5364\relax
\mciteBstWouldAddEndPuncttrue
\mciteSetBstMidEndSepPunct{\mcitedefaultmidpunct}
{\mcitedefaultendpunct}{\mcitedefaultseppunct}\relax
\EndOfBibitem
\bibitem[Cevheri and Yoda(2014)]{cevheriElectrokineticallyDrivenReversible2014}
N.~Cevheri and M.~Yoda, \emph{Lab. Chip}, 2014, \textbf{14}, 1391--1394\relax
\mciteBstWouldAddEndPuncttrue
\mciteSetBstMidEndSepPunct{\mcitedefaultmidpunct}
{\mcitedefaultendpunct}{\mcitedefaultseppunct}\relax
\EndOfBibitem
\bibitem[Yee and Yoda(2018)]{yeeExperimentalObservationsBands2018}
A.~Yee and M.~Yoda, \emph{Microfluid Nanofluid}, 2018, \textbf{22}, 113\relax
\mciteBstWouldAddEndPuncttrue
\mciteSetBstMidEndSepPunct{\mcitedefaultmidpunct}
{\mcitedefaultendpunct}{\mcitedefaultseppunct}\relax
\EndOfBibitem
\bibitem[Lochab \emph{et~al.}(2019)Lochab, Yee, Yoda, Conlisk, and
  Prakash]{lochabDynamicsColloidalParticles2019}
V.~Lochab, A.~Yee, M.~Yoda, A.~T. Conlisk and S.~Prakash, \emph{Microfluid
  Nanofluid}, 2019, \textbf{23}, 134\relax
\mciteBstWouldAddEndPuncttrue
\mciteSetBstMidEndSepPunct{\mcitedefaultmidpunct}
{\mcitedefaultendpunct}{\mcitedefaultseppunct}\relax
\EndOfBibitem
\bibitem[Carlo(2009)]{carloInertialMicrofluidics2009}
D.~D. Carlo, \emph{Lab. Chip}, 2009, \textbf{9}, 3038--3046\relax
\mciteBstWouldAddEndPuncttrue
\mciteSetBstMidEndSepPunct{\mcitedefaultmidpunct}
{\mcitedefaultendpunct}{\mcitedefaultseppunct}\relax
\EndOfBibitem
\bibitem[Huang and Ivory(1999)]{huang_digitally_1999}
Z.~Huang and C.~F. Ivory, \emph{Anal. Chem.}, 1999, \textbf{71},
  1628--1632\relax
\mciteBstWouldAddEndPuncttrue
\mciteSetBstMidEndSepPunct{\mcitedefaultmidpunct}
{\mcitedefaultendpunct}{\mcitedefaultseppunct}\relax
\EndOfBibitem
\bibitem[Stein \emph{et~al.}(2010)Stein, Deurvorst, van~der Heyden, Koopmans,
  Gabel, and Dekker]{stein_electrokinetic_2010}
D.~Stein, Z.~Deurvorst, F.~H.~J. van~der Heyden, W.~J.~A. Koopmans, A.~Gabel
  and C.~Dekker, \emph{Nano Lett.}, 2010, \textbf{10}, 765--772\relax
\mciteBstWouldAddEndPuncttrue
\mciteSetBstMidEndSepPunct{\mcitedefaultmidpunct}
{\mcitedefaultendpunct}{\mcitedefaultseppunct}\relax
\EndOfBibitem
\bibitem[Rempfer \emph{et~al.}(2016)Rempfer, Ehrhardt, Laohakunakorn, Davies,
  Keyser, Holm, and de~Graaf]{Joost_rempfer_selective_2016}
G.~Rempfer, S.~Ehrhardt, N.~Laohakunakorn, G.~B. Davies, U.~F. Keyser, C.~Holm
  and J.~de~Graaf, \emph{Langmuir}, 2016, \textbf{32}, 8525--8532\relax
\mciteBstWouldAddEndPuncttrue
\mciteSetBstMidEndSepPunct{\mcitedefaultmidpunct}
{\mcitedefaultendpunct}{\mcitedefaultseppunct}\relax
\EndOfBibitem
\bibitem[Rempfer \emph{et~al.}(2017)Rempfer, Ehrhardt, Holm, and
  de~Graaf]{rempfer_NP_2017}
G.~Rempfer, S.~Ehrhardt, C.~Holm and J.~de~Graaf, \emph{Macromol. Theory
  Simul.}, 2017, \textbf{26}, 1600051\relax
\mciteBstWouldAddEndPuncttrue
\mciteSetBstMidEndSepPunct{\mcitedefaultmidpunct}
{\mcitedefaultendpunct}{\mcitedefaultseppunct}\relax
\EndOfBibitem
\bibitem[Wanunu \emph{et~al.}(2010)Wanunu, Morrison, Rabin, Grosberg, and
  Meller]{wanunu_electrostatic_2010}
M.~Wanunu, W.~Morrison, Y.~Rabin, A.~Y. Grosberg and A.~Meller, \emph{Nat.
  Nanotechnol.}, 2010, \textbf{5}, 160--165\relax
\mciteBstWouldAddEndPuncttrue
\mciteSetBstMidEndSepPunct{\mcitedefaultmidpunct}
{\mcitedefaultendpunct}{\mcitedefaultseppunct}\relax
\EndOfBibitem
\bibitem[Chou(2009)]{chou_enhancement_2009}
T.~Chou, \emph{J. Chem. Phys.}, 2009, \textbf{131}, 034703\relax
\mciteBstWouldAddEndPuncttrue
\mciteSetBstMidEndSepPunct{\mcitedefaultmidpunct}
{\mcitedefaultendpunct}{\mcitedefaultseppunct}\relax
\EndOfBibitem
\bibitem[Hatlo \emph{et~al.}(2011)Hatlo, Panja, and van
  Roij]{hatlo_Rene_translocation_2011}
M.~M. Hatlo, D.~Panja and R.~van Roij, \emph{Phys. Rev. Lett.}, 2011,
  \textbf{107}, 068101\relax
\mciteBstWouldAddEndPuncttrue
\mciteSetBstMidEndSepPunct{\mcitedefaultmidpunct}
{\mcitedefaultendpunct}{\mcitedefaultseppunct}\relax
\EndOfBibitem
\bibitem[Rabinowitz \emph{et~al.}(2019)Rabinowitz, Edwards, Whittier, Jayant,
  and Shepard]{rabinowitz_nanoscale_2019}
J.~Rabinowitz, M.~A. Edwards, E.~Whittier, K.~Jayant and K.~L. Shepard,
  \emph{J. Phys. Chem. A}, 2019, \textbf{123}, 8285--8293\relax
\mciteBstWouldAddEndPuncttrue
\mciteSetBstMidEndSepPunct{\mcitedefaultmidpunct}
{\mcitedefaultendpunct}{\mcitedefaultseppunct}\relax
\EndOfBibitem
\bibitem[Shin \emph{et~al.}(2017)Shin, Ault, Warren, and
  Stone]{PhysRevX.7.041038}
S.~Shin, J.~T. Ault, P.~B. Warren and H.~A. Stone, \emph{Phys. Rev. X}, 2017,
  \textbf{7}, 041038\relax
\mciteBstWouldAddEndPuncttrue
\mciteSetBstMidEndSepPunct{\mcitedefaultmidpunct}
{\mcitedefaultendpunct}{\mcitedefaultseppunct}\relax
\EndOfBibitem
\bibitem[Zhu \emph{et~al.}(2012)Zhu, Hu, and Xuan]{zhu_electrokinetic_2012}
J.~Zhu, G.~Hu and X.~Xuan, \emph{{Electrophoresis}}, 2012, \textbf{33},
  916--922\relax
\mciteBstWouldAddEndPuncttrue
\mciteSetBstMidEndSepPunct{\mcitedefaultmidpunct}
{\mcitedefaultendpunct}{\mcitedefaultseppunct}\relax
\EndOfBibitem
\bibitem[Patel \emph{et~al.}(2012)Patel, Showers, Vedantam, Tzeng, Qian, and
  Xuan]{patel_microfluidic_2012}
S.~Patel, D.~Showers, P.~Vedantam, T.-R. Tzeng, S.~Qian and X.~Xuan,
  \emph{Biomicrofluidics}, 2012, \textbf{6}, 034102\relax
\mciteBstWouldAddEndPuncttrue
\mciteSetBstMidEndSepPunct{\mcitedefaultmidpunct}
{\mcitedefaultendpunct}{\mcitedefaultseppunct}\relax
\EndOfBibitem
\bibitem[Xuan(2013)]{xuan_reservoir-based_2013}
X.~Xuan, in \emph{Encyclopedia of Microfluidics and Nanofluidics}, ed. D.~Li,
  Springer {US}, 2013, pp. 1--7\relax
\mciteBstWouldAddEndPuncttrue
\mciteSetBstMidEndSepPunct{\mcitedefaultmidpunct}
{\mcitedefaultendpunct}{\mcitedefaultseppunct}\relax
\EndOfBibitem
\bibitem[Patel \emph{et~al.}(2013)Patel, Qian, and
  Xuan]{patel_reservoir-based_2013}
S.~Patel, S.~Qian and X.~Xuan, \emph{{Electrophoresis}}, 2013, \textbf{34},
  961--968\relax
\mciteBstWouldAddEndPuncttrue
\mciteSetBstMidEndSepPunct{\mcitedefaultmidpunct}
{\mcitedefaultendpunct}{\mcitedefaultseppunct}\relax
\EndOfBibitem
\bibitem[Kale \emph{et~al.}(2014)Kale, Patel, Qian, Hu, and
  Xuan]{kale_joule_2014}
A.~Kale, S.~Patel, S.~Qian, G.~Hu and X.~Xuan, \emph{{Electrophoresis}}, 2014,
  \textbf{35}, 721--727\relax
\mciteBstWouldAddEndPuncttrue
\mciteSetBstMidEndSepPunct{\mcitedefaultmidpunct}
{\mcitedefaultendpunct}{\mcitedefaultseppunct}\relax
\EndOfBibitem
\bibitem[Lu \emph{et~al.}(2015)Lu, Dubose, Joo, Qian, and
  Xuan]{lu_viscoelastic_2015}
X.~Lu, J.~Dubose, S.~W. Joo, S.~Qian and X.~Xuan, \emph{Biomicrofluidics},
  2015, \textbf{9}, 014108\relax
\mciteBstWouldAddEndPuncttrue
\mciteSetBstMidEndSepPunct{\mcitedefaultmidpunct}
{\mcitedefaultendpunct}{\mcitedefaultseppunct}\relax
\EndOfBibitem
\bibitem[Kale \emph{et~al.}(2018)Kale, Patel, and
  Xuan]{kale_three-dimensional_2018}
A.~Kale, S.~Patel and X.~Xuan, \emph{Micromachines}, 2018, \textbf{9},
  123\relax
\mciteBstWouldAddEndPuncttrue
\mciteSetBstMidEndSepPunct{\mcitedefaultmidpunct}
{\mcitedefaultendpunct}{\mcitedefaultseppunct}\relax
\EndOfBibitem
\bibitem[Ying \emph{et~al.}(2002)Ying, Bruckbauer, Rothery, Korchev, and
  Klenerman]{klenerman_ying_programmable_2002}
L.~Ying, A.~Bruckbauer, A.~M. Rothery, Y.~E. Korchev and D.~Klenerman,
  \emph{Anal. Chem.}, 2002, \textbf{74}, 1380--1385\relax
\mciteBstWouldAddEndPuncttrue
\mciteSetBstMidEndSepPunct{\mcitedefaultmidpunct}
{\mcitedefaultendpunct}{\mcitedefaultseppunct}\relax
\EndOfBibitem
\bibitem[Ying \emph{et~al.}(2004)Ying, White, Bruckbauer, Meadows, Korchev, and
  Klenerman]{ying_frequency_2004}
L.~Ying, S.~S. White, A.~Bruckbauer, L.~Meadows, Y.~E. Korchev and
  D.~Klenerman, \emph{Biophys. J.}, 2004, \textbf{86}, 1018--1027\relax
\mciteBstWouldAddEndPuncttrue
\mciteSetBstMidEndSepPunct{\mcitedefaultmidpunct}
{\mcitedefaultendpunct}{\mcitedefaultseppunct}\relax
\EndOfBibitem
\bibitem[Harrison \emph{et~al.}(2015)Harrison, Lu, Patel, Thomas, Todd,
  Johnson, Raval, Tzeng, Song, Wang, Li, and
  Xuan]{harrison_electrokinetic_2015}
H.~Harrison, X.~Lu, S.~Patel, C.~Thomas, A.~Todd, M.~Johnson, Y.~Raval, T.-R.
  Tzeng, Y.~Song, J.~Wang, D.~Li and X.~Xuan, \emph{Analyst}, 2015,
  \textbf{140}, 2869--2875\relax
\mciteBstWouldAddEndPuncttrue
\mciteSetBstMidEndSepPunct{\mcitedefaultmidpunct}
{\mcitedefaultendpunct}{\mcitedefaultseppunct}\relax
\EndOfBibitem
\bibitem[Wang \emph{et~al.}(2014)Wang, Bruce, Duch, Patel, Smith, Astier,
  Colgan, Lin, and Stolovitzky]{wang_clog-free_2014}
C.~Wang, R.~Bruce, E.~Duch, J.~Patel, J.~Smith, Y.~Astier, E.~Colgan, Q.~Lin
  and G.~Stolovitzky, 18th International Conference on Miniaturized Systems for
  Chemistry and Life Sciences, MicroTAS 2014, 2014, pp. 1347--1349\relax
\mciteBstWouldAddEndPuncttrue
\mciteSetBstMidEndSepPunct{\mcitedefaultmidpunct}
{\mcitedefaultendpunct}{\mcitedefaultseppunct}\relax
\EndOfBibitem
\bibitem[Wang \emph{et~al.}(2015)Wang, Bruce, Duch, Patel, Smith, Astier,
  Wunsch, Meshram, Galan, Scerbo, Pereira, Wang, Colgan, Lin, and
  Stolovitzky]{wang_hydrodynamics_2015}
C.~Wang, R.~L. Bruce, E.~A. Duch, J.~V. Patel, J.~T. Smith, Y.~Astier, B.~H.
  Wunsch, S.~Meshram, A.~Galan, C.~Scerbo, M.~A. Pereira, D.~Wang, E.~G.
  Colgan, Q.~Lin and G.~Stolovitzky, \emph{ACS Nano}, 2015, \textbf{9},
  1206--1218\relax
\mciteBstWouldAddEndPuncttrue
\mciteSetBstMidEndSepPunct{\mcitedefaultmidpunct}
{\mcitedefaultendpunct}{\mcitedefaultseppunct}\relax
\EndOfBibitem
\bibitem[Karlsen and Dong(2015)]{karlsen_pressure_2015}
H.~Karlsen and T.~Dong, \emph{Chem Eng Commun}, 2015, \textbf{202},
  718--727\relax
\mciteBstWouldAddEndPuncttrue
\mciteSetBstMidEndSepPunct{\mcitedefaultmidpunct}
{\mcitedefaultendpunct}{\mcitedefaultseppunct}\relax
\EndOfBibitem
\bibitem[Zhang \emph{et~al.}(2020)Zhang, Faez, and de~Graaf]{zhu_under_prep}
Z.~Zhang, S.~Faez and J.~de~Graaf, \emph{in preparation}, 2020, \textbf{-},
  --\relax
\mciteBstWouldAddEndPuncttrue
\mciteSetBstMidEndSepPunct{\mcitedefaultmidpunct}
{\mcitedefaultendpunct}{\mcitedefaultseppunct}\relax
\EndOfBibitem
\bibitem[Bocquet and Charlaix(2010)]{bocquet2010nanofluidics}
L.~Bocquet and E.~Charlaix, \emph{Chemical Society Reviews}, 2010, \textbf{39},
  1073--1095\relax
\mciteBstWouldAddEndPuncttrue
\mciteSetBstMidEndSepPunct{\mcitedefaultmidpunct}
{\mcitedefaultendpunct}{\mcitedefaultseppunct}\relax
\EndOfBibitem
\end{mcitethebibliography}

\end{document}